\documentclass[10pt,twocolumn,twoside]{IEEEtran}

\usepackage[cmex10]{amsmath}
\usepackage{amssymb,latexsym}
\usepackage{epsfig}
\usepackage{caption}
\usepackage{color}
\usepackage{setspace}
\usepackage{authblk}
\usepackage{bookmark}
\usepackage{breqn}

\newcommand{\unkum}{\ensuremath{{x}}} 
\newcommand{\nsum}{\ensuremath{{n}}} 
\newcommand{\orgum}{\ensuremath{{y}}} 
\newcommand{\sppum}{\ensuremath{B}}
\newcommand{\umum}{\ensuremath{U_{M B}}}

\newcommand{\esterror}{\ensuremath{\textbf{$\varepsilon$}}} 

\newcommand{\newoperator}[3]{\newcommand*{#1}{\mathop{#2}#3}}

\newoperator{\tr}{\mathrm{tr}}{\nolimits}
\newoperator{\diag}{\mathrm{diag}}{\nolimits}
\newcommand{\expectation}{\ensuremath{\textrm{\textit{E}}}} 
\newcommand{\prob}{\ensuremath{\textrm{\textit{P}}}} 

\newcommand{\trans}{\mathrm{T}}
\newcommand{\conj}{\mathrm{*}}

\newcommand{\myqed}{ \hfill  $\Box$ \\ \vspace{3pt} }

\newtheorem{thm}{\bf{Theorem}}[section]
\newtheorem{cor}{\bf{Corollary}}[section]
\newtheorem{lem}{\bf{Lemma}}[section]

\newtheorem{defn}{\bf{Definition}}[section]
\newtheorem{rem}{\bf{Remark}}[section]
\newtheorem{ex}{\bf{Example}}[section]

\newenvironment{theorem}
{\par\noindent \thm \begin{itshape}\noindent}
{\end{itshape}}

  \newenvironment{lemma}
{\par\noindent \lem \begin{itshape}\noindent}
{\end{itshape}}

  \newenvironment{definition}
{\par\noindent \defn \begin{itshape}\noindent}
{\end{itshape}}

\newenvironment{remark}
{\par\noindent \rem \begin{itshape}\noindent}
{\end{itshape}}

\newenvironment{corollary}
{\par\noindent \cor \begin{itshape}\noindent}
{\end{itshape}\vspace{2pt}} 

\begin{document}

\title{Unitary Precoding and Basis Dependency of MMSE Performance for Gaussian Erasure Channels}

\author{Ay\c ca~\"Oz\c celikkale,
        Serdar~Y\"uksel,
        and~Haldun~M.~Ozaktas 
\thanks{A.~\"Oz\c celikkale is with the Dep. of Signals and Systems, Chalmers University of Technology, SE-41296, 
Gothenburg, Sweden, e-mail: ayca.ozcelikkale@chalmers.se. 
S.~Y\"uksel is with the Dep. of Mathematics and Statistics, Queen's University, K7L3N6, Kingston, Ontario, Canada, e-mail: yuksel@mast.queensu.ca.
H.~M.~Ozaktas is with the Dep. of Electrical Eng., Bilkent University, TR-06800,  Ankara, Turkey, e-mail: haldun@ee.bilkent.edu.tr.
}
\thanks{A.~\"Oz\c celikkale acknowledges the support of  T\" UB\. ITAK  B\.{I}DEB-2211 and B\.{I}DEB-2214. 
 S. Y\"uksel acknowledges the support of the Natural Sciences and Engineering Research Council of Canada (NSERC). 
H. M. Ozaktas acknowledges partial support of the Turkish Academy of Sciences. }
\thanks{Copyright (c) 2014 IEEE. Personal use of this material is permitted.  However, permission to use this material for any other purposes must be obtained from the IEEE by sending a request to pubs-permissions@ieee.org.}
}


 \maketitle


\begin{abstract}
We consider the transmission of a Gaussian vector source over a multi-dimensional Gaussian channel where a random or a fixed subset of the channel outputs are erased. Within the setup where the only encoding operation allowed is a linear unitary transformation on the source, we investigate the MMSE performance, both in average, and also in terms of guarantees that hold with high probability as a function of the system parameters. 
Under the performance criterion of average MMSE, necessary conditions that should be satisfied by the optimal unitary encoders are established and explicit solutions for a class of settings are presented. 
For random sampling of signals that have a low number of degrees of freedom, we present  MMSE bounds that hold with high probability.  
Our results  illustrate how the spread of the eigenvalue distribution  and the unitary transformation contribute to these performance guarantees. The performance of the discrete Fourier transform (DFT) is also investigated.
As a benchmark, we investigate the equidistant sampling of circularly wide-sense stationary (c.w.s.s.) signals, and present the explicit error expression that quantifies the effects of the sampling rate and the eigenvalue distribution of the covariance matrix of the signal. 

These findings may be useful in understanding the geometric dependence of signal uncertainty in a stochastic process. In particular, unlike information theoretic measures such as entropy, we highlight the basis dependence of uncertainty in a signal with another perspective. The unitary encoding space restriction exhibits the most and least favorable signal bases for estimation.
\end{abstract}

\section{Introduction}

We consider the transmission of a Gaussian vector source over a multi-dimensional Gaussian channel where a random or a fixed subset of the channel outputs are erased. We consider the setup where the only encoding operation allowed is a linear unitary transformation on the source. 

\subsection{System Model and Formulation of the Problems} \label{sec:model}
In the following, we present an overview of  the system model  and introduce the family of estimation problems which will be considered in this article.
We first present  a brief description of our problem set-up. We consider the following noisy measurement system
\begin{equation}\label{eqnUnitary:linearSystem}
     \orgum=  {{H}} \unkum+ \nsum = {{H}} U w + \nsum ,
\end{equation}
where $\unkum \in  \mathbb{C}^N$ is the unknown input proper complex Gaussian random
vector, $\nsum \in \mathbb{C}^M$ is the proper complex  Gaussian vector denoting
the measurement noise, and $\orgum \in \mathbb{C}^M$ is the resulting measurement vector. $ {{H}}$ is the $M \times N$ random diagonal sampling matrix.  We assume that $\unkum$ and $\nsum$ are statistically independent zero-mean
random vectors with
covariance matrices ${ K}_{\unkum}=\expectation [\unkum \unkum^\dagger]$, and ${ K}_{\nsum}=\expectation [\nsum \nsum^\dagger]$, respectively. The components of ${ {\nsum}}$ are independent and identically distributed (i.i.d.) with $\expectation [\nsum_i \nsum_i\dagger]=\sigma_n^2>0$.

The unknown signal $\unkum$ comes from the model $\unkum= U w $, where $U$ is a $N \times N$ unitary matrix, and the   components of $w$ are independently (but not necessarily identically) distributed  so that ${K}_{w} =\expectation [w w^\dagger]=\diag(\lambda_1, \dots, \lambda_N)$.
$U$  may be interpreted as the unitary precoder that the signal $w$ is subjected to before going through the channel or the transform that connects the canonical signal domain and the measurement domain.  Hence the singular value decomposition of ${ K}_{\unkum}$ is given by ${ K}_{\unkum}=U {K}_{w} U^\dagger = U \Lambda_x U^\dagger\succeq  0$ where the diagonal matrix denoting the eigenvalue distribution of the covariance matrix of $\unkum$ is given by  $\Lambda_x={K}_{w}=\diag(\lambda_1, \dots, \lambda_N)$.
We are interested in the minimum mean-square error (MMSE) associated with estimating  $\unkum$  (or equivalently $w$), that is   $\expectation[ ||x - \expectation [x| y ]||^2 = \expectation[ ||w - \expectation [w| y ]||^2 $.
Throughout the article, we assume that the receiver has access to channel realization information, i.e. the realization of the random sampling matrix $H$.

We interpret the eigenvalue distribution of ${ K}_{\unkum}$ as a measure of the low dimensionality of the signal. The case where most of the eigenvalues are zero and the nonzero eigenvalues have equal values is interpreted as the counterpart of the standard, exactly sparse signal model  in compressive sensing. The case where most of the power of the signal is carried by a few eigenvalues,  is interpreted to model the more general signal family which has an {\it{effectively}} low degree of freedom.
Yet, we note that our model is different from the classical compressive sensing setting.  
Here we assume that the receiver knows the covariance matrix $K_x$, i.e.  it has full knowledge of the support of the input.

Our investigations can be summarized under two main problems. In the first problem, we search for the best unitary encoder under the performance criterion of average (over random sampling matrix $H$) MMSE.

\vspace{5pt}
{{\textrm{{\bf{Problem P1}} (Best Unitary Encoder For Random Channels):}}
Let $\mathbb{U^N}$ be the set of $N \times N$ unitary matrices: $\{U \in \mathbb{C}^N :U^\dagger U=I_N\}$. We consider the following minimization problem
\begin{equation}
 \inf_{U \in {\mathbb{U}^{N}}}  \expectation_{H} \left[ \expectation_{S} [ ||x - \expectation [x| y ]||^2 ] \right],
\end{equation}
where the expectation with respect to the random measurement matrix and the expectation with respect to random signals involved  is denoted by $\expectation_{H}[.]$, and $\expectation_{S}[.]$, respectively.
\vspace{5pt}

In the second avenue, we will regard the MMSE performance as a random variable and consider performance guarantees that hold with high probability with respect to random sampling matrix $H$.
We will not explicitly cast this problem as an optimal unitary precoding problem as we have done in Problem P1. Nevertheless, the results will illustrate the favorable transforms through the coherence parameter  $\mu=\max_{i,j}|u_{ij}|$, which is extensively used in the compressive sensing literature \cite{candesRomberg_2007,Tropp_2008,DonohoHuo_2001}.

\vspace{5pt}
 {\textrm{{\bf{Problem P2}} (Error Bounds That Hold With High Probability):}}  Let $\tr(K_x)=P$.  Let  $D(\delta)$ be the smallest number satisfying $\sum_{i=1}^D \lambda_i \geq \delta P$, where $\delta \in (0,1]$ and  $\lambda_1 \geq \lambda_2, \dots, \geq \lambda_N$ . Assume that  the effective number of  degrees of freedom of the signal is small, so that there exists a $D(\delta)$ small compared to $N$ with $\delta$ close to $1$.
We investigate  nontrivial lower bounds (i.e. bounds close to 1) on
\begin{equation}
\prob \bigg(  \expectation_{S}[ ||x - \expectation [x| y ]||^2 ] < f_{P2}(\Lambda_x,U,\sigma_n^2) \bigg)
\end{equation}
for some function $f_{P2}(.)$ which  denotes a sufficiently small error level given total power of the unknown signal, $\tr{(K_x)}$, and
 the noise level $\sigma_n^2$.
\vspace{5pt}

\subsection{Literature Review and Main Contributions}

In the following, we provide a brief overview of the related literature. In this article, we consider the Gaussian erasure channel, where each component of the unknown vector is erased independently and with equal probability, and the transmitted components are observed through Gaussian noise. This type of model may be used to formulate various types of transmission with low reliability scenarios, for example Gaussian channel with impulsive noise \cite{Tulino_2007, Tulino_2007preprint}.
This measurement model is also related to the measurement scenario typically considered in the compressive sensing framework \cite{CandesRomberg_2006,CandesRombergTao_2006} under which each component is  erased independently and with equal probability.
The only difference between these two models is the explicit inclusion of the noise in the former. 
In this respect, our work contributes to  the understanding of the MMSE performance of such measurement schemes under noise. 
Although there are compressive sensing studies that consider scenarios where the signal recovery is done by explicitly acknowledging the presence of noise, a substantial amount of the work focuses on the noise-free scenario.
A particularly relevant exception is \cite{TulinoCaireVerduShamai_2013}, where the authors work on the same setting as the one in our article with Gaussian inputs. This work considers the scenario under which the signal support is not known whereas we assume that the signal support is known at the receiver.

The problem of optimization of precoders or input covariance matrices is formulated in literature under different performance criteria:  When the channel is not random, \cite{Basar_1980}  considers a related trace minimization problem, and  \cite{Witsenhausen_1975} a determinant maximization problem, which,   in our formulation, correspond to optimization of the MMSE and mutual information performance, respectively.   \cite{YuRheeBoydCioffi_2004}, \cite{PerezCruzRodriguesVerdu_2010}  formulate the problem with the criterion of mutual information, whereas \cite{LeePetersen_1976} focuses on the MMSE and \cite{YangRoy_1994} on determinant of the mean-square error matrix. \cite{Palomar_2003,Palomar_2004} present a general framework based on Schur-convexity.  In these works the channel is known at the transmitter, hence it is possible to shape the  input according to the channel. When the channel is a Rayleigh or Rician fading channel, \cite{Kashyap_2003} investigates the best linear encoding problem without restricting the encoder to be unitary.
\cite{Telatar_1999} focuses on the problem of maximizing the mutual information for a Rayleigh fading channel.  \cite{Tulino_2007}, \cite{Tulino_2007preprint} consider the erasure channel as in our setting, but with the aim of maximizing the ergodic capacity.
Optimization of linear precoders are also utilized in communications applications, for instance in broadcasting of video over wireless networks where each user operates under a different channel quality \cite{JakubczakKatabi_2010}.

In Section~\ref{secUnitary:probFixedDomain} and Section~\ref{secUnitary:probProjections}, we investigate how the results in random matrix theory mostly presented in compressive sampling framework can be used to find bounds on  the MMSE associated with the described measurement scenarios. We note that there are studies that consider the MMSE in compressive sensing framework such as \cite{TulinoCaireVerduShamai_2013,RanganFletcherGoyal_2012,EladYavneh_2009, ProtterYavnehElad_2010}, 
which focus on the scenario where the receiver does not know the location of the signal support (eigenvalue distribution). In our case we assume that the receiver has full knowledge of the signal covariance matrix, hence the signal support.

\vspace{5pt}
{\bf Contributions of the paper.} In view of the above literature review, our main contributions can be summarized as follows:
We formulate the problem of finding the most favourable unitary transform under average (over random sampling) MMSE criterion (Problem P1).  We investigate the convexity properties of this optimization problem, obtain necessary conditions of optimality through variational equalities, and solve some special cases. Among these we have identified special cases where DFT-like unitary transforms (unitary transforms with  $|u_{ij}|^2=\frac{1}{N}$) are optimal coordinate transforms. We also show that, in general, DFT is not the optimal unitary transform.
For the noiseless case, we have also observed that the identity transform turns out to be universally the worst unitary transform regardless of the eigenvalue decomposition.

On Problem 2, under the assumption of known signal support, our results quantify the error associated with estimating a signal with effectively low degree of freedom from randomly selected samples, in the $\ell_2$ framework of MMSE estimation instead of  the $\ell_1$ framework of typical compressive sensing results. The performance guarantees for signals that have strictly low degree of freedom follows from recent  random matrix theory results in a straightforward manner.   We present MMSE performance guarantees that illustrate the trade-off between the eigenvalue distribution of the covariance matrix of the signal (effective number of degrees of freedom) and the unitary transform (spread of the uncertainty in the channel). Although there are a number of works in compressive sensing literature that consider signals with low effective degree of freedom (see for instance \cite[Sec 2.3]{Rauhut_2010}, and the references therein) our findings do not directly follow from these results. 
As a benchmark,  we investigate the case where $U$ is the DFT matrix and the sampling is done equidistantly.   In this case, the covariance matrix is circulant, and the resulting signal $x$ is referred as circularly wide-sense stationary, which is a natural way to model wide-sense stationary signals in finite dimension. We present the explicit MMSE expression in this case.  Although this result comes from simple linear algebra arguments, to the best of our knowledge they do not appear elsewhere in the literature.

Our results show that the general form of error bounds that hold with high probability are the same with the error expression associated with the equidistant sampling of band pass c.w.s.s. signals, but with a lower effective SNR term. The loss in the effective SNR may be interpreted to come through  two multiplicative loss factors, one due to random sampling, (which is present even when all the insignificant eigenvalues are zero), and the other due to the presence of nonzero insignificant eigenvalues.

\subsection{Motivation}\label{sec:motivation}

Our motivation for studying these problems, in particular our focus on the best unitary precoders, is two-fold.

In the first front, we would like to characterize the impact of the unitary precoder on estimation performance, since  such restrictions occur in both physical contexts and applications.  Optimization of linear precoders or input covariance matrices arises naturally in many signal estimation and communication applications including transmission over multiple input multiple output (MIMO)  channels, for instance with unitary precoders \cite{KimParkLoveKim_2009, LoveHeath_2005}. Our restriction of the transformation matrix to a unitary transformation rather than a more general matrix (say a  noiselet transform) is motivated by some possible restrictions in the measurement scenarios and the potential numerical benefits of unitary transforms. In many measurement scenarios one may not be able to pass the signal through an arbitrary transform before random sampling, and may have to measure it just after it passes through a unitary transform. Using more general transforms may cause additional complexity  or may not be feasible. Possible scenarios where unitary transformations play an important role can be given in the context of optics: The propagation of light is governed by a diffraction integral, a convenient approximation of which is the Fresnel integral, which constitutes a unitary transformation on the input field (see, for instance \cite{ozaktas_book}). Moreover, a broad class of optical systems  involving arbitrary concatenations of lenses, mirrors, sections of free space, quadratic graded-index media, and phase-only spatial light modulators can be well represented by unitary transformations \cite{ozaktas_book}. Hence if one wants to estimate the light field by measuring the field after it propagates in free space or passes through such a system, one has to deal with a unitary transform, but not a more general one.
Furthermore, due to their structure, unitary transforms have low complexity numerical implementations. For instance, the DFT which is among the most favourable transforms for high probability results is also very attractive from numerical point of view, since  there is a fast algorithm with complexity $N \log(N)$ for taking the DFT of a signal.

Our second, and primary motivation for our work comes from the desire to understand the geometry of statistical dependence in random signals. We note that the dependence of signal uncertainty in the signal basis has been considered in different contexts in the information theory literature. The concepts that are traditionally used in the information theory literature as measures of dependency or uncertainty in signals (such as  the number of degrees of freedom, or the entropy) are mostly defined independent of the coordinate system in which the signal is to be measured.  As an example one may consider the Gaussian case: the entropy solely depends on the eigenvalue spectrum of the covariance matrix, hence making the concept blind to the coordinate system in which the signal lies in.
On the other hand, the approach of applying coordinate transformations to orthogonalize signal components is adopted in many signal reconstruction and information theory problems. For example the rate-distortion function for a Gaussian random vector is obtained by applying an uncorrelating transform to the source, or approaches such as the Karhunen-Lo\'eve expansion are used extensively. Also, the compressive sensing community heavily makes use of the notion of coherence of bases, see for example \cite{candesRomberg_2007,Tropp_2008,DonohoHuo_2001}. The coherence of two bases, say the  intrinsic signal domain $\psi$ and the orthogonal measurement system   $\phi$  is measured with  $\mu=\max_{i,j}|u_{ij}|$, $U=  \phi \psi$ providing a measure of how concentrated the columns of $U$ are. When $\mu$ is small, one says the mutual coherence is small.  As the coherence gets smaller,  fewer  samples  are required to provide good performance guarantees.

Our study of the measurement problems in this article confirms that signal recovery performance depends  substantially on total uncertainty of the signal (as measured by the differential entropy); but also illustrates that the basis plays an important role in the measurement problem. The total uncertainty in the signal as quantified by information theoretic measures such as entropy (or eigenvalues) and the spread of this uncertainty (basis) reflect different aspects of the dependence in a signal. Our framework makes it possible to study these relationships in a systematic way, where the eigenvalues of the covariance matrix provide a well-defined measure of uncertainty.   Our analysis here illustrates the interplay between these two concepts.

Before leaving this section, we would like to discuss the role of DFT-like transforms in our setting.
In Problem P2 we will see that, in terms of the sufficiency conditions stated, DFT-like unitary matrices will provide the most favorable performance guarantees, in the sense that fixing the bound on the probability of error, they will require the least number of measurements.
We also note the following: In compressive sensing literature, the performance results depend on some constants, and  it is reported in \cite[Sec. 4.2]{Rauhut_2010} that better constants are available for the DFT matrix.  Moreover, for the DFT matrix, it is known that the technical condition that states the nonzero entries of the signal has a random sign pattern which is typical of such results can be removed  \cite[Sec. 4.2]{Rauhut_2010}. \footnote{We note that there are some recent results that suggest that the results obtained by the DFT matrix may be duplicated for Haar distributed unitary matrices: limiting distributions of eigenvalues of  Haar distributed unitary matrices and the DFT matrix behave similarly under random projections, see for instance \cite{Farrell_2011}, and the eigenvalues of certain sums (for instance, ones like in the MMSE expression) involving  Haar distributed unitary matrices can be obtained  from the eigenvalues of individual components and are well-behaved \cite{TulinoCaireVerduShamai_2013,TulinoVerdu_2004}. } Hence the current state of art in compressive sensing suggests the idea that the DFT is the most favorable unitary transform for such random sampling scenarios. Yet, we will see that for Problem P1, DFT is not, in general an optimal encoder within the class of unitary encoders.

\subsection{Preliminaries and Notation}\label{sec:prelim}
In the following, we present a few definitions and notations that will be used throughout the article.
Let $\tr{(K_x)}=P$. Let $D(\delta)$ be the smallest number satisfying $\sum_{i=1}^D \lambda_i \geq \delta P$, where $\delta \in (0,1]$. Hence for $\delta$ close to one, $D(\delta)$  can be considered as an effective rank of the covariance matrix and also the effective number of ``degrees of freedom" (DOF) of the signal family. For $\delta$ close to one, we drop the dependence on $\delta$ and use the term effective DOF  to represent $D(\delta)$. A closely related concept is the (effective) bandwidth. We use the term ``bandwidth" for the DOF of a signal family whose canonical domain is the Fourier domain, i.e. whose unitary transform is given by the DFT matrix.

The transpose, complex conjugate and  complex conjugate transpose of a matrix $A$ is denoted by $A^\trans$, $A^\conj$ and $A^\dagger$, respectively. 
The $t^{th}$ row $k^{th}$ column entry of $A$ is denoted by $a_{tk}$. 
The eigenvalues of a matrix $A$  are denoted in decreasing order as $\lambda_1(A) \geq \lambda_2(A), \dots, \geq \lambda_N(A)$.

Let $\sqrt{-1}=j$.  The entries of the $N \times N$ DFT matrix  are given by $v_{tk}={1 \over \sqrt N} e^{j  {2 \pi \over N} t k}$,
where $0 \leq t \, , k \leq N-1$. We note that the DFT matrix is the diagonalizing unitary transform for all circulant matrices \cite{GrayToeplitzReview}.  In general, a circulant matrix  is determined by its first row and defined by the relationship  $C_{tk}=C_{0\, \text{mod}_N (k-t)}$,  where rows and columns are indexed  by $t$ and $k$,  $0 \leq t \, , k \leq N-1$, respectively.

We now review the expressions for the MMSE estimation. Under a given measurement matrix $H$, by standard arguments the MMSE estimate is given by  $\expectation [x | y ]=\hat{x}=K_{xy}  {K_{y}}^{-1} y$, where $K_{xy}=E[{xy^\dagger}]=K_x H^\dagger$, and $K_{y}=E[{yy^\dagger}]=H K_x H^\dagger + K_n $. We note that since $K_n \succ 0$, we have $K_y \succ 0$, and hence  $K_y^{-1}$ exists.  The associated MMSE can be expressed as \cite[Ch2]{b_andersonMoore_optFiltering}
\begin{dgroup}
\begin{dmath}
 \expectation_{S} [ ||x - \expectation [x | y ]||^2 ]
  = \tr(K_x-K_{xy} K_y^{-1} K_{xy} ^\dagger) 
  \end{dmath}
  \begin{dmath}\label{eqUnitary:errorlong}
   = \tr(K_x) - \tr(K_x H^\dagger (H K_x H^\dagger + K_n)^{-1} H K_x)  
   \end{dmath}
   \begin{dmath}
    = \tr(U \Lambda_x U^\dagger) - \tr(U \Lambda_x U^\dagger H^\dagger (H U \Lambda_x U^\dagger H^\dagger + K_n)^{-1} H U \Lambda_x U^\dagger)
    \end{dmath}
\end{dgroup}
 Let $B=\{i : \lambda_i>0 \}$, and let $U_B$ denote the $N \times |B|$ matrix formed by taking the columns of $U$ indexed by $B$. Similarly, let $\Lambda_{x,B}$ denote the $|B| \times |B|$ matrix by taking the columns and rows of  $\Lambda_{x}$ indexed by $B$ in the respective order. We note that $U_B^\dagger U_B=I_{|B|}$, whereas the equality $U_B U_B^\dagger=I_N$ is not true unless $|B|=N$. Also note that $\Lambda_{x,B}$ is always invertible. The singular value decomposition of $K_x$ can be written as $K_x= U \Lambda_x U^\dagger=U_B \Lambda_{x,B} U_B^\dagger$. Hence the error may be rewritten as
 \begin{dgroup}
\begin{dmath*}
  \expectation_{S} [ ||x - \expectation [x | y ]||^2 ] \\
  \end{dmath*}
  \begin{dmath*}
    = \tr(U_B \Lambda_{x,B} U_B^\dagger)-\tr(U_B \Lambda_{x,B} U_B^\dagger H^\dagger (H U_B \Lambda_{x,B} U_B^\dagger H^\dagger + K_n)^{-1} HU_B \Lambda_{x,B} U_B^\dagger)
    \end{dmath*}
    \begin{dmath}
     \label{eqUnitary:tr}
     = \tr( \Lambda_{x,B}) -\tr( \Lambda_{x,B} U_B^\dagger H^\dagger (H U_B \Lambda_{x,B} U_B^\dagger H^\dagger + K_n)^{-1} HU_B \Lambda_{x,B} )
     \end{dmath}
     \begin{dmath}
     \label{eqUnitary:Woodburry}
     =  \tr{((\Lambda_{x,B}^{-1}+\frac{1}{\sigma_n^2} U_B^{\dagger} H ^{\dagger} H U_B)^{-1})}
     \end{dmath}
\end{dgroup}
where \eqref{eqUnitary:tr} follows from the identity $\tr(U_B M U_B^\dagger)=\tr( M U_B^\dagger U_B)=\tr( M)$ with an arbitrary  matrix $M$ with consistent dimensions. Here \eqref{eqUnitary:Woodburry} follows from the fact that $\Lambda_{x,B}$ and $K_n$ are nonsingular and the Sherman-Morrison-Woodbury identity, which has the following form for our case (see for example \cite{HendersonSearle_1981} and the references therein)
\begin{align*}
K_1 -  K_1 A^\dagger  (A K_1 A^\dagger+K_2)^{-1} & A K_1\\
    &= (K_1^{-1}+ A^\dagger K_2^{-1} A)^{-1},
\end{align*}
where $K_1$ and $K_2$ are nonsingular.

Here is a brief summary of the rest of the article: In Section~\ref{secUnitary:avg_erasureChannel}, we formulate the problem of finding the most favorable unitary transform under average MMSE criterion (Problem P1).
In Section~\ref{secUnitary:prob}, we find performance guarantees for the MMSE estimation that hold with high probability (Problem P2). Our benchmark case for the high probability results, the error associated with the equidistant sampling of circularly wide-sense stationary signals, is presented in Section~\ref{secUnitary:cwss}.
We conclude in Section~\ref{secUnitary:conc}.


\section{Average MMSE} \label{secUnitary:avg_erasureChannel} 

In this section, we investigate the optimal unitary precoding problem with the performance criterion of  average (with respect to random sampling matrix $H$)  MMSE. 
In Section~\ref{secUnitary:prob}, we will focus on MMSE guarantees that hold with high probability (w.r.t. $H$).

We assume that the receiver knows the channel information, whereas the transmitter only knows the channel probability distribution.
We consider the following measurement strategies: a) ({\it{Random Scalar Gaussian Channel:}})  $H=e_i^T$, $i=1,\ldots, N$ with probability $1 \over N$, where $e_i \in \mathbb{R}^N$ is the $i^{th}$ unit vector. We denote this sampling strategy with $S_s$. b) ({\it{Gaussian Erasure Channel}})  $H= diag(\delta_i)$, where $\delta_i$ are i.i.d. Bernoulli random variables with probability of success $p \in [0,1]$.  We denote this sampling strategy with $S_b$.

Let $\mathbb{U^N}$ be the set of $N \times N$ unitary matrices: $\{U \in \mathbb{C}^N :U^\dagger U=I\}$.
We consider the following minimization problem
\begin{equation} \label{eqn:main:unitary}
 \inf_{U \in {\mathbb{U}^{N}}}  \expectation_{H} \left[ \expectation_{S} [ ||x - \expectation [x| y ]||^2 ] \right],
\end{equation}
where the expectation with respect to $H$ is over admissible measurement strategies $S_s$ or $S_b$. Hence we want to determine the best unitary encoder for the random scalar Gaussian channel or  Gaussian erasure channel.
\newline

We note that  \cite{Tulino_2007} and \cite{Tulino_2007preprint} consider the erasure channel model ($S_b$ in our notation) with the aim of maximizing the ergodic capacity. Their formulations let the transmitter also shape the eigenvalue distribution of the source, whereas ours does not.

We note that by solving \eqref{eqn:main:unitary} for the measurement scheme in \eqref{eqnUnitary:linearSystem}, one also obtains the solution for the  generalized the set-up  $y=H V x+n$, where $V$ is any unitary matrix: Let $U_o$ denote an optimal unitary matrix for the scheme in \eqref{eqnUnitary:linearSystem}. Then $V^\dagger U_o \in  {\mathbb{U}^{N}}$ is an optimal unitary matrix for the generalized set-up. 

\subsection{First order necessary conditions for optimality}
Here we discuss the convexity properties of the optimization problem and give the first order necessary conditions for optimality.  We note that we do not utilize these conditions for finding the optimal unitary matrices. The reader not interested in these results can directly continue on to Section~\ref{secUnitary:specialCases}.

Let the possible sampling schemes be indexed by the variable $k$, where $1 \leq k \leq N$ for $S_s$, and $1 \leq k \leq 2^N$ for $S_b$. Let $H_k$ be the corresponding sampling matrix. Let $p_k$ be the probability of the $k^{th}$ sampling scheme.

We can express the objective function as follows
\begin{dgroup*}
\begin{dmath*}
 \expectation_{H,S} [ ||x - \expectation [x | y ]||^2 ] 
 \end{dmath*}
 \begin{dmath*}
 =  \expectation_{H} [ \tr{((\Lambda_{x,B}^{-1}+\frac{1}{\sigma_n^2} U_B^{\dagger} H ^{\dagger} H U_B)^{-1}})  ]
 \end{dmath*}
 \begin{dmath}
 \label{eqnUnitary:errorprob}
 = \sum_k p_k  \tr{((\Lambda_{x,B}^{-1}+\frac{1}{\sigma_n^2} U_B^{\dagger} H_k ^{\dagger} H_k U_B)^{-1})}.
 \end{dmath}
\end{dgroup*}
The objective function is a continuous function of $U_B$. We also note that the feasible set defined by
 $\{U_B \in \mathbb{C}^{N \times |B|} :U_B^\dagger U_B=I_{|B|}\}$  is a closed and bounded subset of $\mathbb{C}^n$, hence compact. Hence the minimum is attained since we are minimizing a continuous function over a compact set (but the optimum $U_B$ is not necessarily unique).

We note that in general, the feasible region is not a convex set. Let $U_1, U_2 \in \mathbb{U^N}$  and $\theta \in [0,1]$. In general $\theta U_1 + (1-\theta) U_2 \notin  \mathbb{U^N}$. For instance let $N=1$, $U_1=1$,  $U_2=-1$, $\theta U_1 + (1-\theta) U_2 =2 \theta -1 \notin \mathbb{U}^1,  \quad \forall \, \theta \in [0,1]$.
Even if the unitary matrix constraint is relaxed, we observe that the objective function is in general neither a convex or a concave function of the matrix $U_B$.  To see this, one can check  the second derivative to see if $\nabla_{U_B}^2  f(U_B) \succeq 0$ or $ \nabla_{U_B}^2  f(U_B)  \preceq 0$, where $f(U_B)= \sum_k p_k  \tr{((\Lambda_{x,B}^{-1}+\frac{1}{\sigma_n^2} U_B^{\dagger} H_k ^{\dagger} H_k U_B)^{-1})}$. For example, let $N=1$, $U \in \mathbb{R}$, $\sigma_n^2=1$, $\lambda>0$, and $p>0$ for $S_b$. Then $f(U)= \sum_{k} p_k \frac{1}{\lambda^{-1}+ U^{\dagger} H_k ^{\dagger} H_k U}$ can be written as $f(U)= (1-q)  \lambda +q \frac{1}{\lambda^{-1}+ U^{\dagger}  U}$, where $q \in (0,1]$ is the probability that the one  possible measurement is done. That is $q=1$ for $S_s$, and  $q=p$ for $S_b$.  Hence  $\nabla_{U}^2  f(U)=  q \,  2  \, \frac{3  U^2 -\lambda^{-1} }{(\lambda^{-1}+ U^2)^3}$, whose sign changes depending on $\lambda$, and $U$. Hence neither $\nabla_{U}^2  f(U) \succeq 0$ nor $ \nabla_{U}^2  f(U) \preceq 0$ holds for all $U \in \mathbb{R}$.

In general,  the  objective function depends only on $U_B$, not $U$. 
If $U_B$ satisfying  $U_B^\dagger U_B=I_{|B|}$, with $|B|<N$ is an optimal solution, then a properly chosen set of column(s)  can be added to $U_B$ so that a unitary matrix $U$ is formed. Any such $U$ will have the same objective value with $U_B$, and hence will also be an optimal solution. 
Therefore it is sufficient to consider the constraint   $\{U_B:U_B^\dagger U_B=I_{|B|}\}$, instead of the condition $\{U:U^\dagger U=I_N\}$,  while optimizing the objective function. We also note that if  $U_B$ is an optimal solution, $\exp(j \theta) U_B$ is also an optimal solution, where  $0 \leq \theta \leq 2\pi$.

Let $u_i$ be the $i^{th}$ column of $U_B$. We can write the unitary matrix constraint as follows:
\begin{eqnarray}
u_i^\dagger u_k=
\begin{cases}
1,  & \mbox{if } i=k, \\
0, & \mbox{if } i\neq k.
\end{cases}
\end{eqnarray}
with $i=1,\ldots,|B|$, $k=1,\ldots,|B|$. Since $u_i^\dagger u_k = 0$, iff  $u_k^\dagger u_i=0$, it is sufficient to consider $k \leq i$.  Hence this constraint may be rewritten as
\begin{eqnarray} \label{eqUnitary:unitaryconst}
e_i^\trans (U_B^\dagger U_B-I_{|B|})e_k=0, 
\end{eqnarray}
 with $i=1,\ldots,|B|,$    $ k=1,\ldots,i$. Here $e_i \in \mathbb{R}^{|B|}$ is the $i^{th}$ unit vector.

We note that constraint gradients (gradients of the conditions in \eqref{eqUnitary:unitaryconst}) are linearly independent for any matrix $U_B$ satisying $U_B^\dagger U_B=I_B$ \cite{ayca_phd}.  Hence the linear independence constraint qualification (LICQ) holds for any feasible $U_B$ \cite[Defn.12.4]{nocedalWright_numericalOptimization}. Therefore, the first order condition $\nabla_{U_B}  L( U_B,\nu, \upsilon) =0$ together with the condition  $U_B^\dagger U_B=I_B$ is necessary for optimality \cite[Thm 12.1]{nocedalWright_numericalOptimization}, where  $ L( U_B,\nu, \upsilon) $ is the Lagrangian for some Lagrangian multiplier vectors $\nu$, and $\upsilon$.
The Lagrangian can be expressed as follows 
\begin{align}
\nonumber
 L( U_B,\nu, \upsilon) & =
 \sum_k  p_k  \tr{((\Lambda_{x,B}^{-1}+ \frac{1}{\sigma_n^2} U_B^{\dagger} H_k ^{\dagger} H_k U_B)^{-1})} \\
 \nonumber
  & +\sum_{ (i,k) \in \bar{\gamma}} \nu_{i,k} e_i^\trans (U_B^\dagger U_B-I_{|B|}) e_k \\
  \nonumber
  & +\sum_{ (i,k) \in \bar{\gamma}} \nu_{i,k}^\conj e_i^\trans (U_B^\trans U_B^\conj-I_{|B|})e_k \\
 \label{eqnUnitary:lagrangian} 
  & +\sum_{ k=1}^{|B|} \upsilon_{k} e_k^\trans (U_B^\dagger U_B-I_{|B|}) e_k, 
\end{align}
where $\nu_{i,k} \in \mathbb{C}$, ${ (i,k) \in \bar{\gamma}}$ and  $\upsilon_k \in \mathbb{R}$,  $k \in \{1, \ldots,|B|\}$ are the Lagrange multipliers. Here $\bar{\gamma}$ is defined as the following set of pairs of indices $\bar{\gamma} =\{(i,k)|  i=1,\ldots,|B|, \,\,    k=1,\ldots,i-1\}$.

The first order necessary condition $\nabla_{U_B}  L( U_B,\nu, \upsilon) =0$ can be expressed more explicitly as follows:
\begin{lemma}
The following condition is necessary for optimality
\begin{align}
\nonumber
& \sum_k p_k {(\Lambda_{x,B}^{-1}+\frac{1}{\sigma_n^2} U_B^{\dagger} H_k ^{\dagger} H_k U_B)^{-2}}  U_B^\dagger H_k^\dagger H_k \\
\nonumber
& = \sum_{ (i,k) \in \bar{\gamma}} \nu_{i,k} e_k e_i^\trans U_B^\dagger +\sum_{ (i,k) \in \bar{\gamma}} \nu_{i,k}^* e_i e_k^\trans U_B^\dagger \\
& +\sum_{ k=1}^{|B|}\upsilon_{k} e_k  e_k^\trans U_B^\dagger,
\end{align}
with $\nu_{i,k}$ and  $\upsilon_k$ Lagrange multipliers as defined above, taking possibly different values. 
\end{lemma}\\

{\bf{Proof:}} The proof is based on the guidelines for  optimization problems and derivative operations involving complex variables presented in  \cite{Brandwood_1983,HjorungnesGesbert_2007,b_magnusNeudecker_matrixCalculus}.  Please see \cite{ayca_phd} for the complete proof.

\begin{remark}
For $S_s$, we can analytically show that this condition is satisfied by the DFT matrix and the identity matrix. It is not surprising that both the DFT matrix and the identity matrix satisfy these equations, since this optimality condition is the same for both minimizing and maximizing the objective function.  We show that the DFT matrix is indeed one of the possibly many minimizers for the case where the values of the nonzero eigenvalues are equal in Lemma~\ref{lemUnitary:scalarEigFlat}.  The maximizing property of the  identity matrix in the noiseless case is investigated in  Lemma~\ref{lemUnitary:worstCoordinateTransformation}.

In Section~\ref{secUnitary:prob}, we show that with the DFT matrix, the MMSE is small with high probability for signals that have small number of degrees of freedom.
Although these observations and the other special cases presented in Section~\ref{secUnitary:specialCases} may suggest  the result that the DFT matrix may be an optimum solution for the general case,   we show that this is not the case by presenting a counterexample where another unitary matrix not satisfying $|u_{ij}|^2=1/N$ outperforms the DFT [Lemma \ref{lemUnitary:DFTnotoptimal}].
\end{remark}

 \subsection{Special cases} \label{secUnitary:specialCases}
In this section,  we consider some related special cases.  For random scalar Gaussian channel, we will show that when the nonzero eigenvalues are equal  any covariance matrix (with the given eigenvalues) having a constant diagonal is an optimum solution [Lemma~\ref{lemUnitary:scalarEigFlat}]. This includes  Toeplitz covariance matrices or covariance matrices with any unitary transform satisfying $|u_{ij}|^2=1/N$. We note that the DFT matrix satisfies $|u_{ij}|^2=1/N$ condition, and always produces circulant covariance matrices. 
We will also show that for both channel structures, for the noiseless case (under some conditions) regardless of the entropy or the number of degrees of freedom of a signal, the worst coordinate transformation is the same, and given by the identity matrix [Lemma~\ref{lemUnitary:worstCoordinateTransformation}].

For the general Gaussian erasure channel model,  we will show that when only one of the eigenvalues is nonzero (i.e. rank of the covariance matrix is one), any unitary transform satisfying $|u_{ij}|^2=1/N$ is an optimizer [Lemma~\ref{lemUnitary:DOF1DFToptimum}]. We will also show that under the relaxed condition $\tr(K_x^{-1})=R$,  the best covariance matrix is circulant, hence the best unitary transform is the DFT matrix [Lemma~\ref{lemUnitary:GECtraceinverse}].
We note that Ref.~\cite{Tulino_2007preprint} proves the same result under the aim of maximizing mutual information with a power constraint on $K_x$, i.e. $\tr{(K_x)} \leq P$. Ref.~\cite{Tulino_2007preprint} further finds the optimal eigenvalue distribution, whereas in our case, the condition on the trace of the inverse is introduced as a relaxation, and in the original problem we are interested, the eigenvalue distribution is fixed.

In the next section, we will show that the observations presented in  compressive sensing literature implies that the MMSE is small with high probability when $|u_{ij}|^2=1/N$. Although all these observations may suggest the result that the DFT matrix may be an optimum solution in the general case, we will show that this is not the case by presenting a counterexample where another unitary matrix not satisfying $|u_{ij}|^2=1/N$ outperforms the DFT matrix [Lemma~\ref{lemUnitary:DFTnotoptimal}].

Before moving on, we note the following relationship between the eigenvalue distribution and the MMSE.
Let $H \in \mathbb{R}^{M \times N}$ be a sampling matrix formed by taking $1 \leq 3 M \leq N$ rows from the  identity matrix. Assume that $\Lambda_x \succ 0$. Let the eigenvalues of a matrix $A$  be denoted in  decreasing order as $\lambda_1(A) \geq \lambda_2(A), \dots, \geq \lambda_N(A)$. The MMSE can be  expressed as follows \eqref{eqUnitary:Woodburry}
\begin{dgroup}
 \begin{dmath}
 \expectation [ ||x - \expectation [x | y ]||^2 ]  = \tr{((\Lambda_{x}^{-1}+\frac{1}{\sigma_n^2} U^{\dagger} H ^{\dagger} H U)^{-1})}  
   \end{dmath}
 \begin{dmath}
= \sum_{i=1}^N \frac{1}{\lambda_i(\Lambda_x^{-1}  + \frac{1}{\sigma_n^2}U^{\dagger}  H^\dagger H U  )} 
  \end{dmath}
 \begin{dmath}
 = \sum_{i=M+1}^{N} \frac{1}{\lambda_i(\Lambda_{x}^{-1} + \frac{1}{\sigma_n^2}U^{\dagger}  H^\dagger H U )} +\sum_{i=1}^{M}  \frac{1}{\lambda_i(\Lambda_x^{-1}  + \frac{1}{\sigma_n^2}U^{\dagger}  H^\dagger H U)} 
   \end{dmath}
 \begin{dmath}
 \label{eqnUnitary:lowerBound_1}
  \geq  \sum_{i=M+1}^{N} \frac{1}{\lambda_{i-M}(\Lambda_x^{-1} )}+\sum_{i=1}^{M}  \frac{1}{\lambda_i(\Lambda_x^{-1}  + \frac{1}{\sigma_n^2}U^{\dagger}  H^\dagger H U)}    
    \end{dmath}
   \begin{dmath}
  \label{eqnUnitary:lowerBound_2}
    \geq  \sum_{i=M+1}^{N} \frac{1}{\lambda_{i-M}(\Lambda_x^{-1} )}+\sum_{i=1}^{M}  \frac{1}{\frac{1}{\lambda_{N-i+1}(\Lambda_x)}+ \frac{1}{\sigma_n^2}}  
      \end{dmath}
     \begin{dmath}
=  \sum_{i=M+1}^{N} \lambda_{N-i+M+1}(\Lambda_x)+\sum_{i=N-M+i}^{N}  \frac{1}{\frac{1}{\lambda_{i}(\Lambda_x)}+ \frac{1}{\sigma_n^2}} \\
   \end{dmath}
 \begin{dmath}
  \label{eqUnitary:lowerBound}
  =  \sum_{i=M+1}^{N} \lambda_{i}(\Lambda_x)+\sum_{i=N-M+1}^{N}  \frac{1}{\frac{1}{\lambda_{i}(\Lambda_x)}+ \frac{1}{\sigma_n^2}},
  \end{dmath}
\end{dgroup}
where we have used case (b)  of Lemma~\ref{lemUnitary:rankeigHorn} in \eqref{eqnUnitary:lowerBound_1}, and the fact that $\lambda_i(\Lambda_x^{-1}  + \frac{1}{\sigma^2} U^{\dagger}  H^\dagger H U) \leq  \lambda_i(\Lambda_x^{-1} )+ \frac{1}{\sigma^2}\lambda_{1}(U^{\dagger}  H^\dagger H U)= \lambda_i(\Lambda_x^{-1} )+ \frac{1}{\sigma^2}$ in \eqref{eqnUnitary:lowerBound_2}.

\begin{lemma} [4.3.3, 4.3.6, \cite{Horn_matrix}]\label{lemUnitary:rankeigHorn} Let $A_1,A_2 \in \mathbb{C}^{N\times N}$ be Hermitian matrices. (a) Let  $A_2$ be positive semi-definite. Then  $\lambda_{i}(A_1+A_2) \geq \lambda_{i}(A_1)$,  $i=1, \ldots, N.$  (b) Let the rank of $A_2$ be at most $M$, $ 3 M \leq N$.  Then $\lambda_{i+M}(A_1+A_2) \leq \lambda_{i}(A_1)$,  $i=1, \ldots, N-M.$
\end{lemma}

This lower bound in \eqref{eqUnitary:lowerBound}  is consistent with our intuition: If the eigenvalues are well-spread, that is $D(\delta)$ is large in comparison  to $N$ for $\delta$ close to 1, the error cannot be made small without making a large number of measurements. The first term in \eqref{eqUnitary:lowerBound} may be obtained by the following intuitively appealing alternative argument: The energy compaction property of Karhunen-Lo\`eve expansion guarantees that 
 the best representation of this signal with $M$ variables in mean-square error sense is obtained by first decorrelating the signal with $U^\dagger$ and then using the random variables that correspond to the highest $M$ eigenvalues. The mean-square error of such a representation is given by the sum of the  remaining eigenvalues, i.e.     $\sum_{i=M+1}^{N} \lambda_{i}(\Lambda_x)$.
Here we make measurements before decorrelating the signal, and each component is measured with noise. Hence the error of our measurement scheme is lower bounded by the error of the optimum scheme, which is exactly the first term in  \eqref{eqUnitary:lowerBound}.  The second term is the MMSE associated with the measurement scheme in which $M$ independent variables with variances given by the  $M$ smallest eigenvalues of $\Lambda_x$  are observed through  i.i.d. noise.

\begin{lemma}  \label{lemUnitary:scalarEigFlat} {\it{[Scalar Channel: Eigenvalue Distribution Flat]}} Let $\tr(K_x)=P$. Assume that  the nonzero eigenvalues are equal, i.e. $\Lambda_{x,B}=\frac{P}{|B|} I_B$. Then the minimum average error for $S_s$ is given by
\begin{equation}
P-\frac{P}{|B|}+\frac{1}{1+\frac{P}{N} \frac{1}{\sigma_n^2}}\frac{P}{|B|},
\end{equation}
which is achieved by covariance matrices with constant diagonal. In particular, covariance matrices whose unitary transform is the DFT matrix satisfy this property.
\end{lemma}

{\bf{Proof:}}
(Note that if none of the eigenvalues are zero, $K_x=I$ regardless of the unitary transform, hence the objective function value does not depend on it.)
The objective function may be expressed as \eqref{eqnUnitary:errorprob}
\begin{dgroup*}
\begin{dmath*}
 \expectation_{H,S} [ ||x - \expectation [x | y ]||^2 ] 
 \end{dmath*}
 \begin{dmath*}
 = \sum_{k=1}^N \frac{1}{N} \tr{( \frac{|B|}{P} I_B+ \frac{1}{\sigma_n^2} U_B^{\dagger} H_k ^{\dagger} H_k U_B)^{-1}}  
 \end{dmath*}
 \begin{dmath}
 \label{eqUnitary:order}
 = \frac{P}{|B|}  \sum_{k=1}^N  \frac{1}{N} (|B|-1+ {(1+  \frac{P}{|B|} \frac{1}{\sigma_n^2} H_k U_B U_B^{\dagger} H_k
^{\dagger} )^{-1}}) 
\end{dmath}
\begin{dmath*}
=   \frac{P}{|B|}  (|B|-1)+\sum_{k=1}^N    \frac{P}{|B|} \frac{1}{N} {(1+  \frac{P}{|B|} \frac{1}{\sigma_n^2} e_k^\dagger U_B U_B^{\dagger} e_k  )^{-1}},
\end{dmath*}
\end{dgroup*}
where in $\eqref{eqUnitary:order}$ we have used Lemma 2 of \cite{Kashyap_2003}.
We now consider the minimization of the following function
\begin{align}
\nonumber
 \sum_{k=1}^N {(1+   \frac{P}{|B|} \frac{1}{\sigma_n^2}  e_k^\dagger U_B U_B^{\dagger} e_k  )^{-1}}  
 &=   \sum_{k=1}^N  \frac{1}{1+   \frac{P}{|B|} \frac{1}{\sigma_n^2} \frac{|B|}{P}  z_k} \\
 \label{eqn:zk}
 &  =  \sum_{k=1}^N  \frac{1}{1+ \frac{1}{\sigma_n^2} z_k},
\end{align}
where  $(U_B U_B^{\dagger})_{kk}=\frac{|B|}{P}  (K_x)_{kk}=\frac{|B|}{P}  z_k$ with $z_k= (K_x)_{kk}$. Here $z_k\geq 0$ and  $\sum_k z_k=P$, since $\tr{(K_x)}=P$. We note that the goal is the minimization of  a convex function over a convex region.
 We note that the function in \eqref{eqn:zk} is a Schur-convex function of $z_k$'s. This follows from, for instance, Prop. C1 of  \cite[Ch.~3]{b_MarshallOlkin} and the fact that ${1}/{(1+ ({1}/{\sigma_n^2}) z_k)}$ is convex. Together with the power constraint, this reveals that the optimum $z_k$ is given by $z_k = P/N$.
We observe that this condition is equivalent to require that the covariance matrix has constant diagonal.  This condition can be always satisfied; for example with a Toeplitz covariance matrix or with any unitary transform satisfying $|u_{ij}|^2=1/N$. We note that the DFT matrix satisfies $|u_{ij}|^2=1/N$ condition, and always produces circulant covariance matrices. \myqed

\begin{lemma}  \label{lemUnitary:worstCoordinateTransformation}
{ [\it{Worst Coordinate Transformation}]}
We now consider the random scalar channel $S_s$ without noise, and consider the following maximization problem which searches for the worst coordinate system for a signal to lie  in: 
\begin{equation}
 \sup_{U \in \mathbb{U^N}}  \expectation [  \sum_{t=1}^N  [||x_t - \expectation [x_t | y ]||^2] ],
\end{equation}
where $y= x_i$ with probability  $\frac{1}{N}$, $ i=1,\ldots,N$ and $\tr(K_x)=P$.

The solution to this problem is as follows: The maximum value of the objective function is $P-P/N$.  $U=I$ achieves this maximum value.
\end{lemma}

\begin{remark}
We emphasize that this result does not depend on the eigenvalue spectrum $\Lambda_x$.
\end{remark}

\begin{remark}
We note that when some of the eigenvalues of the covariance matrix are identically zero, the eigenvectors  corresponding to the zero eigenvalues can be chosen freely (of course as long as the resulting transform $U$ is unitary).
\end{remark}

{\bf{Proof:}}
 The objective function may be written as
 \begin{align}
 \nonumber
& \expectation  [  \sum_{t=1}^N [||x_t   -  \expectation [x_t | y]||^2] ]  \\
 & =  {1 \over N} \sum_{i=1}^N  \sum_{t=1}^N  \expectation [||x_t - \expectation [x_t |x_{i} ]||^2] ] \\
 & = {1 \over N} \sum_{i=1}^N  \sum_{t=1}^N (1-\rho_{i,t}^{2})\sigma_{x_t}^{2},
\end{align}
where $\rho_{i,t}=\frac{\expectation [x_{t} x_{i}^\dagger]}{(\expectation [||x_{t}||^{2}]\expectation [\,||x_{i}||^{2}])^{1/2}}$ is the correlation coefficient between $x_{t}$ and $x_{i}$, assuming  $\sigma_{x_t}^{2} =\expectation [||x_{t}||^{2}]>0$,  $\sigma_{x_i}^{2}>0$. (Otherwise one may set $\, \rho_{i,t}=1$ if $i=t$, and $\rho_{i,t}=0$ if $i \neq j$.) Now we observe that $\sigma_{t}^{2} \geq 0$, and  $0 \leq |\rho_{i,t}|^{2} \leq 1$. Hence the maximum value of this function is given by $\rho_{i,t}=0,\,\, \forall\, t,i \,\, \text{s.t.}\,\, t \neq i$. We observe that any diagonal unitary matrix $U = \diag(u_{ii})$, $|u_{ii}|=1$ (and also any $\bar{U} =U \Pi$, where $\Pi$ is a permutation matrix) achieves this maximum value. In particular, the identity transform $U=I_N$ is an optimal solution. 

We note that a similar result holds for $S_b$:  Let  $y= H x$.  The optimal value of $ \sup_{U \in {\mathbb{U}^{N}}}  \expectation_{H,S} [ ||x - \expectation [x| y ]||^2 ],$
where the expectation with respect to $H$ is over  $S_b$ is $(1-p) \tr{(K_x)}$, which is achieved by any $U \Pi$,  $U = \diag(u_{ii})$, $|u_{ii}|=1$, $\Pi$ is a permutation matrix. \myqed

\begin{lemma}  \label{lemUnitary:DOF1DFToptimum} 
{\it{[Rank 1 Covariance Matrix]}}
Suppose $|B|=1$, i.e.  $\lambda_k=P>0$, and $\lambda_j=0$, $j \neq k, j \in 1,\ldots, N$. The minimum error under $S_b$ is given by the following expression
\begin{equation}
\expectation{[\frac{1}{\frac{1}{P}+ \frac{1}{\sigma_n^2} \frac{1}{N} \sum_{i=1}^N \delta_i }]},
\end{equation}
where this optimum is achieved by any unitary matrix whose $k^{th}$ column entries satisfy $|u_{ik}|^2=1/N$, $i=1,\ldots, N$.
\end{lemma}

{\bf{Proof:}}
 Let $v={[v_1, \ldots,v_n ]}^{\mathrm{T}}$,  $v_i=|u_{ki}|^2$, $i=1,\ldots,N$, where $\mathrm{T}$ denotes transpose. We note the following
\begin{align}
\nonumber
& \expectation[{\tr{(\frac{1}{P}+ \frac{1}{\sigma_n^2} U_B^\dagger H^\dagger H U_B)^{-1}}}] \\
&= \expectation[{\frac{1}{\frac{1}{P}+ \frac{1}{\sigma_n^2} \sum_{i=1}^N \delta_i |u_{ki}|^2}}]\\
&=  \expectation[{\frac{1}{\frac{1}{P}+\frac{1}{\sigma_n^2} \sum_{i=1}^N \delta_i v_i}}].
\end{align}
The proof uses an argument in the proof of  \cite[Thm. 1]{Telatar_1999}, which is also used in \cite{Kashyap_2003}. Let  $\Pi_i \in \mathbb{R}^{N\times N}$ denote the permutation matrix indexed by $i=1,\ldots, N!$. We note that a feasible vector $v$ satisfies $\sum_{i=1}^N v_i=1$, $v_i \geq 0$, which forms a convex set.   We observe that for any such $v$, weighted sum of all permutations of $v$, $\bar{v}=\frac{1}{N!}  \sum_{i=1}^{N!} \Pi_i v = (\frac{1}{N} \sum_{i=1}^N v_i ) [1,\ldots,1]^T = [ \frac{1}{N},\ldots, \frac{1}{N}]^T \in \mathbb{R}^N$ is a constant vector  and also feasible. We note that $g(v)=\expectation[{\frac{1}{\frac{1}{P} + \frac{1}{\sigma_n^2}\sum_i \delta_i v_i}}]$ is a convex function of $v$ over the feasible set. Hence $g(v) \geq g(\bar{v})=g([1/N,\ldots, 1/N])$ for all $v$, and $\bar{v}$ is the optimum solution. Since there exists a unitary matrix satisfying $|u_{ik}|^2=1/N$ for any given $k$ (such as any unitary matrix whose $k^{th}$ column is any column of the DFT matrix), the claim is proved.\myqed

\begin{lemma}  \label{lemUnitary:GECtraceinverse}  
{\it{[Trace constraint on the inverse of the covariance matrix]}}
Let $K_x^{-1} \succ 0 $. Instead of fixing the eigenvalue distribution, let us consider the relaxed constraint $\tr(K_x^{-1})=R$. Let $K_n \succ 0 $.  Then an optimum solution for 
\begin{align}
&\arg \min_{K_x^{-1}} \expectation_{H,S} [ ||x - \expectation [x | y ]||^2 ] \\
\nonumber
&= \arg \min_{K_x^{-1}} \expectation_{H} [(\tr(K_x^{-1}+\frac{1}{\sigma_n^2} H^\dagger K_n^{-1} H)^{-1}]
\end{align}
under $S_b$  is a circulant matrix.
\end{lemma}

{\bf{Proof:}}
The proof uses an argument in the proof of \cite[Thm. 12]{Tulino_2007preprint}, \cite{Tulino_2007}.
Let  $\Pi$ be the following permutation matrix,
\begin{eqnarray}
 \Pi= \left[\begin{array}{cccc}0 & 1 & \cdots & 0 \\0 & 0 & 1  & 0 \cdots \\  \vdots &  & \ddots & \vdots \\1 & \cdots &  0 & 0\end{array}\right].
\end{eqnarray}
We observe that $\Pi$ and $\Pi^l$ ($l^{th}$ power of $\Pi$) are unitary matrices.
We form the following matrix $\bar{K}_{x}^{-1}= \frac{1}{N} \sum_{l=0}^{N-1}\Pi^l K_{x}^{-1} (\Pi^l)^{\dagger}$,
which also satisfies the power constraint $\tr{(\bar{K}_{x}^{-1})}=R$. We note that since $K_{x}^{-1} \succ 0$,  so is $\bar{K}_{x}^{-1} \succ 0$, hence  $\bar{K}_{x}^{-1}$ is well-defined.
\begin{align}
\nonumber
&\expectation  \left[ \tr \left( (\frac{1}{N} \sum_{l=0}^{N-1}\Pi^l K_{x}^{-1} (\Pi^l)^{\dagger}+\frac{1}{\sigma_n^2}  H^\dagger K_n^{-1} H)^{-1} \right) \right]  \\
\label{eqn:Jensencirc}
&\leq   \frac{1}{N} \sum_{l=0}^{N-1}  \expectation \left[ \tr \left( (\Pi^l K_{x}^{-1} (\Pi^l)^{\dagger}+\frac{1}{\sigma_n^2} H^\dagger K_n^{-1} H)^{-1} \right) \right]    \\ 
\label{eqnUnitary:trinv}   
& =   \frac{1}{N} \sum_{l=0}^{N-1}  \expectation \left[\tr \left( (K_{x}^{-1} + \frac{1}{\sigma_n^2}  (\Pi^l)^{\dagger}  H^\dagger K_n^{-1} H \Pi^l )^{-1} \right) \right]   \\
\label{eqnUnitary:samedist} 
& =   \frac{1}{N} \sum_{l=0}^{N-1}  \expectation \left[\tr \left( (K_{x}^{-1} + \frac{1}{\sigma_n^2} H^\dagger K_n^{-1} H )^{-1} \right) \right]    \\
& =  \expectation \left[\tr \left(  (K_{x}^{-1} + \frac{1}{\sigma_n^2} H^\dagger K_n^{-1} H )^{-1} \right)\right] 
\end{align}
We note that  $\tr(({M+K_n^{-1}})^{-1})$ is a convex function of  $M$ over the set $M \succ 0$, since $\tr(M^{-1})$ is a convex function (see for example \cite[Exercise 3.18]{boyd}), and composition with an affine mapping preserves convexity  \cite[Sec. 3.2.2]{boyd}. Hence \eqref{eqn:Jensencirc} follows from Jensen's Inequality applied to the summation forming $\bar{K}_{x}^{-1}$.  \eqref{eqnUnitary:trinv} is due to the fact that $\Pi^l$s are unitary and trace is invariant under unitary transforms.  \eqref{eqnUnitary:samedist} follows from the fact that $H \Pi^l$  has the same distribution with $H$.
Hence we have shown that $\bar{K}_{x}^{-1}$ provides a lower bound for arbitrary ${K}_{x}^{-1}$ satisfying the power constraint. Since $\bar{K}_{x}^{-1}$  is circulant and  also satisfies the power constraint $\tr{(\bar{K}_{x}^{-1})}=R$, an optimum ${K}_{x}^{-1}$ is also circulant. \myqed

We note that we cannot follow the same argument for the  constraint $\tr(K_x)=P$, since the objective function is concave in $K_x$ over  the set  $K_x \succ 0$. This can be seen as follows: The error can be expressed as  $\expectation [ ||x - \expectation [x | y ]||^2 ] =\tr{(K_e)}$, where  $K_e=K_x-K_{xy} K_y^{-1} K_{xy}^\dagger$. We note that $K_e$ is the Schur complement of $K_y$ in $K= [K_y\,\, K_{yx}; K_{xy}\,\, K_x]$, where $K_y=H K_x H^\dagger+K_n$, $K_{xy}=K_x H^\dagger$. Schur complement is matrix concave in $K \succ 0$,  for example see \cite[Exercise 3.58]{boyd}.  Since trace is a linear operator, $\tr(K_e)$ is concave in $K$. Since $K$ is an affine mapping of $K_x$, and composition with an affine mapping preserves concavity \cite[Sec. 3.2.2]{boyd}, $\tr(K_e)$ is concave in $K_x$.

\begin{lemma}  \label{lemUnitary:DFTnotoptimal}
{ [DFT is not always optimal]}
The DFT matrix is, in general, not an optimizer of the minimization problem stated in \eqref{eqn:main:unitary} for the Gaussian erasure channel.
\end{lemma}

{\bf{Proof:}}
We provide a counterexample to prove the claim of the lemma: An example where a unitary matrix not satisfying $|u_{ij}|^2=1/N$ outperforms the DFT matrix. Let $N=3$. Let $\Lambda_x=\diag(1/6,2/6,3/6)$, and $K_n=I$. Let $U$ be
 \begin{equation}
 U_0=\left[\begin{array}{ccc}1/\sqrt{2} & 0 & 1/\sqrt{2} \\0 & 1 & 0 \\-1/\sqrt{2} & 0 & 1/\sqrt{2}\end{array}\right]
\end{equation}
 Hence $K_x$ becomes
  \begin{equation}
 K_x=\left[\begin{array}{ccc}1/3 & 0 & 1/6 \\0 & 1/3 & 0 \\1/6 & 0 & 1/3\end{array}\right]
 \end{equation}
We write the average error  as a sum conditioned on the number of measurements as $J(U)=\sum_{M=0}^3 p^M (1-p)^{3-M} e_M(U)$, where $e_M$ denotes the total error of all cases where $M$ measurements are done. Let $e(U)=[e_0(U), e_1(U), e_2(U), e_3(U) ]$. The calculations reveal that  $e(U_0)=[1, 65/24, 409/168, 61/84]$  whereas $e(F)=[1, 65/24, 465/191, 61/84]$, where $F$ is the DFT matrix. We see that all the entries are the same with the DFT case, except $e_2(U_0) < e_2(F)$, where  $e_2(U_0)=409/168 \approx 2.434524$  and $e_2(F)=465/191 \approx 2.434555$. Hence $U_0$ outperforms the DFT matrix.

We note that our argument covers any unitary matrix that is formed by changing the order of the columns of the DFT matrix, i.e. any matching of the given eigenvalues and the columns of the DFT matrix: $U_0$ provides better performance than any $K_x$ formed by using the given eigenvalues and any unitary matrix formed with columns from the DFT matrix.  \myqed

\section{MMSE Bounds That Hold With High Probability} \label{secUnitary:prob}
In this section, we focus on MMSE bounds that hold with high probability. 
As a preliminary work, we will first consider a sampling scenario which will serve as a benchmark  in the subsequent sections: estimation of a c.w.s.s. signal from its equidistant samples. Circularly wide-sense stationary signals provide a natural analogue for stationary signals in the finite dimension, hence in a sense they are the most basic signal type one can consider in a sampling setting.  Equidistant sampling strategy is the sampling strategy which one commonly employs in a sampling scenario. Therefore, the error associated with equidistant sampling under c.w.s.s. model forms an immediate candidate for comparing the error bounds associated with random sampling scenarios. 

\subsection{Equidistant Sampling of Circularly Wide-Sense Stationary Random Vectors} \label{secUnitary:cwss}
In this section, we consider the case where  $x$ is a  zero-mean, proper, c.w.s.s. Gaussian random vector. 
Hence the covariance matrix of  $x$ is circulant, and the unitary transform $U$ is fixed, and given by the DFT matrix by definition \cite{GrayToeplitzReview}.


We assume that the sampling is done equidistantly: Every $1$ out of $\Delta N$ samples are taken. We let $M={N \over \Delta N} \in \mathbb{Z}$, and assume that the first component of the signal is measured, for convenience. 

By definition, the eigenvectors of the covariance matrix is given by the columns of the DFT matrix, where the elements of k$^{th}$ eigenvector is given by  $u_{tk}={1 \over \sqrt N} e^{j  {2 \pi \over N} t k}$, $0 \leq t \leq N-1$.  We denote the associated eigenvalue with $\lambda_k$, $0 \leq k \leq N-1$ instead of indexing the eigenvalues  in decreasing order.

\begin{lemma} \label{lemUnitary:mmse_cwssnoisy} 
The MMSE of estimating $x$ from the equidistant noisy samples $y$ as described above is given by the following expression
\begin{align}
 \label{eqnUnitary:mmse_cwssmainnoisy}
 &\expectation [ ||x - \expectation [x | y ]||^2 ]  \\ 
 \nonumber
 &=   \sum_{k=0}^{M-1} ( \sum_{i=0}^{\Delta N-1} \lambda_{i M +k} -\sum_{i=0}^{\Delta N-1}  \frac{\lambda_{i M +k}^2}{\sum_{l=0}^{\Delta N-1} (\lambda_{l M +k}+\sigma_n^2)})
 \end{align} 
\end{lemma}

{\bf{Proof:}} 
Proof is provided in Section~\ref{sec:pf:lemUnitary:mmsecwss}.

A particularly important special case is the error associated with the estimation of a  band-pass signal: 
 
\begin{corollary}\label{lemUnitary:mmse_cwssnoisybandlimited} 
Let $\tr(K_x) =P$. Let the eigenvalues be given as  $\, \lambda_i =  \frac{P}{|B|},\, \text{if}  \,\, 0 \leq i  \leq |B|-1 $, and  $\lambda_i = 0,\, \text{if}  \,\,  |B| \leq i \leq N-1$. If $M \geq |B|$, then the error can be expressed as follows
\begin{align} \label{eqnUnitary:mmse_cwssnoisy_bandlimited}
 \expectation [ ||x - \expectation [x | y ]||^2 ]  =  \frac{1}{1+ \frac{1}{{\sigma^2_n}} \frac{P}{|B|}  \frac{M}{N}} P
 \end{align} 
\end{corollary}
We note that this expression is of  the form $\frac{1}{1+\text{SNR}} P$, where $\text{SNR}= \frac{1}{{\sigma^2_n}} \frac{P}{|B|}  \frac{M}{N}$. This expression will serve as a benchmark in the subsequent sections. 


\subsection{Flat Support} \label{secUnitary:probFixedDomain}
We now focus on MMSE bounds that hold with high probability. 
In this section, we assume that all nonzero eigenvalues are equal, i.e. $\Lambda_{x,B}=\frac{P} {|B|} I_{|B|}$, where $|B| \leq N$ . We will consider more general eigenvalue distributions in Section~\ref{secUnitary:probProjections}. We present bounds on the MMSE depending on the support size and the number of measurements that hold with high probability. 
These results illustrate how the results in matrix theory mostly presented in compressive sampling framework can provide MMSE bounds.
%
We note that the problem we tackle here is inherently different from the $\ell_1$ set-up considered in traditional compressive sensing problems. Here we consider the problem of estimating a Gaussian signal in  Gaussian noise under the assumption the support is known. It is known that the best estimator in this case is the linear MMSE estimator. On the other hand, in scenarios where one refers to $\ell_1$ characterization, one typically does not know the support of the signal.
We note that there are studies that consider the unknown support scenario in a MMSE framework, such as \cite{TulinoCaireVerduShamai_2013,RanganFletcherGoyal_2012,EladYavneh_2009, ProtterYavnehElad_2010}.

We consider the set-up in \eqref{eqnUnitary:linearSystem}. The random sampling operation is modelled with a  $M \times N$  sampling matrix $H$, whose rows are taken from the identity matrix as dictated by the sampling operation. We let $ \umum =H U_B$ be the $M \times |\sppum|$ submatrix of $U$ formed by taking $|\sppum|$ columns  and $M$ rows as dictated by $B$ and $H$, respectively.
The MMSE can be expressed as follows \eqref{eqUnitary:Woodburry}
\begin{align}
\nonumber
 & \expectation_S [ ||x - \expectation [x | y ]||^2 ] \\
 \nonumber
    &=  \tr{((\Lambda_{x,B}^{-1}+\frac{1}{\sigma_n^2} U_B^{\dagger} H ^{\dagger} H U_B)^{-1})}  \\
  \nonumber
    &=  \sum_{i=1}^{|B|}   \frac{1}{\lambda_i{(\frac{|B|}{P} I_B+\frac{1}{\sigma_n^2} \umum ^{\dagger} \umum)} } \\
   \label{eqnUnitary:noisyerr}
     &=   \sum_{i=1}^{|B|} \frac{1}{\frac{|B|}{P} +\frac{1}{\sigma^2_n} \lambda_{i} ({ \umum}^\dagger  \umum)}.
\end{align}
We see that the estimation error is determined by the eigenvalues of the matrix $\umum^\dagger  \umum$.
We note that many results in compressive sampling framework make  use of the bounds on the eigenvalues of this matrix. We now use one of these results to bound the MMSE performance.
The discussion here may not be surprising for readers who are familiar with the tools used in the compressive sensing community, since the analysis here is related to recovery problems with high probability. However, this discussion highlights how these results are mimicked with the MMSE criterion and how the eigenvalues of the covariance matrix can be interpreted as measure of low effective degree of freedom of a signal family. We note that different eigenvalue bounds in the literature can be used, we pick one of these bounds from the literature to make the constants explicit.

\begin{lemma} \label{lemUnitary:randomMeasFixedSupport} Let $U$ be an $N \times N$ unitary matrix with  $\sqrt{N} \max_{k,j} |u_{k,j}| = \mu(U)$. Let the signal have fixed support $\sppum$ on the signal domain. Let the sampling locations be chosen uniformly at random from the set of all subsets of the given size $M$, $M \leq N$. Let noisy measurements with noise power $\sigma^2_n$  be done at these  $M$ locations.
Then for sufficiently large $M(\mu)$, the error is bounded from above with high probability:
\begin{equation} \label{eqnUnitary:randomMeasFixedSupport} 
 \expectation_S [ ||x - \expectation [x | y ]||^2 ] <   \frac{1}{ 1+\frac{1}{\sigma^2_n}\frac{{0.5} M}{N} \frac{P}{|B|}} P
\end{equation}
More precisely, if
\begin{equation}\label{eqnUnitary:candesRombergCond}
M \geq |\sppum| \mu^2(U) \max(C_1 \log |\sppum|, C_2 \log (3/\delta))
\end{equation}
for some positive constants $C_1$ and $C_2$, then
\begin{equation} \label{eqnUnitary:candesRombergError}
\prob ( \expectation_S [ ||x - \expectation [x | y ]||^2 ] \geq  \frac{1}{ 1+\frac{1}{\sigma^2_n}\frac{{0.5} M}{N} \frac{P}{|B|}} P) \leq \delta.
\end{equation}
In particular, when the measurements are noiseless, the error is zero with probability at least $1-\delta$.
\end{lemma}

{\bf{Proof}:}
We first note that $\| { \umum}^\dagger  \umum-I \| <  c$ implies $1-c < \lambda_i({ \umum}^\dagger  \umum) < 1+c$. Consider Theorem 1.2 of \cite{candesRomberg_2007}.  Suppose that  $M$ and $|\sppum|$ satisfies \eqref{eqnUnitary:candesRombergCond}.  Now looking at  Theorem 1.2, and noting  the scaling of the  matrix $U^\dagger U =N I$ in  \cite{candesRomberg_2007}, we see that  ${ P} (0.5\frac{M}{N} < \lambda_i({ \umum}^\dagger  \umum) < 1.5\frac{M}{N}) \geq 1-\delta$. By \eqref{eqnUnitary:noisyerr} the result follows.

For the noiseless measurements case, let $\esterror= \expectation_S [ ||x - \expectation [x | y ]||^2 ]$, and $A_{\sigma^2_n}$ be the event $\{\esterror <   {\sigma^2_n}     \frac{|B|}{ {\sigma^2_n} \frac{|B|}{P} +\frac{{0.5} M}{N}}\}$
Hence
\begin{eqnarray}
 \lim_{\sigma^2_n \rightarrow 0} \prob (A_{\sigma^2_n}) &=& \lim_{\sigma^2_n \rightarrow 0} \expectation [1_{A_{\sigma^2_n}}] \\
 &=& \expectation [ \lim_{\sigma^2_n \rightarrow 0} 1_{A_{\sigma^2_n}}] \\
       &=&   \prob (\esterror = 0)
\end{eqnarray}
where we have used Dominated Convergence Theorem to change the order of the expectation and the limit. By \eqref{eqnUnitary:candesRombergError} $\prob (A_{\sigma^2_n}) \geq 1- \delta$, hence  $\prob (\esterror = 0) \geq 1- \delta $.  We also note that in the noiseless case, it is enough to have $\lambda_{\min}( \umum^\dagger  \umum)$ bounded away from zero to have zero error with high probability, the exact value of the bound is not important. \myqed 


We note that when the other parameters are fixed, as $\max_{k,j} |u_{k,j}| $ gets smaller, fewer number of samples are required. Since  $\sqrt{1/N} \leq  \max_{k,j} |u_{k,j}| \leq  1$ , the unitary transforms that provide the most favorable guarantees are the ones satisfying $|u_{k,j}|=\sqrt{1/N} $. We note that for any such unitary transform,  the covariance matrix has constant diagonal with $(K_x)_{ii}=P/N$ regardless of the eigenvalue distribution.
Hence with any measurement scheme with $M$, $M \leq N$ noiseless measurements,
the reduction in the uncertainty is guaranteed to be at least proportional to the number of measurements, i.e. the error satisfies $\esterror \leq P-\frac{M}{N} P$.

\begin{remark}
We note that the coherence parameter $\mu(U)$ takes the largest value possible for the DFT: $\mu(U)=\sqrt{N} \max_{k,j} |u_{k,j}| =1$. Hence due to the role of  $\mu(U)$ in the error bounds, in particular in the conditions of the lemma (see \eqref{eqnUnitary:candesRombergCond}), the DFT may be interpreted as one of the most favorable unitary transforms possible in terms of the sufficiency conditions stated. We recall that for a c.w.s.s. source, the unitary transform associated with the covariance matrix is given by the DFT.
Hence we can conclude that  Lemma~\ref{lemUnitary:randomMeasFixedSupport} is applicable to these signals. That is,  among signals with a covariance matrix with  a given rectangular eigenvalue spread,  c.w.s.s. signals are among the ones that can be estimated with low values of error with high probability with a given number of randomly located measurements.
\end{remark}

We finally note that using the argument employed in Lemma~\ref{lemUnitary:randomMeasFixedSupport}, one can also find MMSE  bounds for the adverse scenario where a signal  with random support is sampled at fixed locations. (We will still assume that the receiver has access to the support set information.)
In this case the results that explore the bounds on the eigenvalues of random submatrices obtained by uniform column sampling,  such as Theorem 12 of \cite{Tropp_2008} or Theorem 3.1 of \cite{Darses_2011}, can be used in order to bound the estimation error.

\subsubsection{Discussion}
We now compare the error bound found above with the error associated with equidistant sampling of a  low pass circularly wide-sense stationary  source. 
We consider the special case where $x$ is a band pass signal with  $\lambda_0 = \dots =\lambda_{|B|-1} = P/|B|$,  $\lambda_{|B|}=\ldots=\lambda_{N-1}=0$.  
By Corollary~\ref{lemUnitary:mmse_cwssnoisybandlimited}, if  the number of measurements $M$ is larger than the bandwidth, that is $M \geq |B|$,  the error associated with the equidistant sampling scheme can be expressed as 
\begin{equation} \label{eqnUnitary:randomMeasFixedSupport:disc}
 \expectation [ ||x - \expectation [x| y ]||^2 ] =  \frac{1}{ 1 +\frac{P}{|B|} \frac{1}{{\sigma^2_n}} \frac{M}{N} } \, P.
\end{equation}
Comparing  \eqref{eqnUnitary:randomMeasFixedSupport} with this expression, we observe the following: The expressions are of the same general form, $\frac{1}{1+ c \, \text{SNR}} P$, where $\text{SNR} \triangleq  \frac{P}{|B|} \frac{1}{{\sigma^2_n}} \frac{M}{N}$, with $0 \leq c \leq 1$ taking different values for different cases.
We also note that in \eqref{eqnUnitary:randomMeasFixedSupport}, the choice of $c=0.5$, which is the constant chosen for the  eigenvalue bounds in \cite{candesRomberg_2007}, is for convenience. It could have been chosen  differently by choosing a different probability $\delta$ in \eqref{eqnUnitary:candesRombergError}.
We also observe that effective $\text{SNR}$ takes its maximum value with $c=1$ for the deterministic equidistant sampling strategy corresponding to the minimum error value among these two expressions. 
In random sampling case, $c$ can only take smaller values, resulting in larger and hence worse error bounds. 
We note that one can choose $c$ values closer to 1, but then the probability these error bounds hold decreases, that is better error bounds can be obtained at the expense of lower degrees of guarantees that these results will hold.

The result of Lemma~\ref{lemUnitary:mmse_cwssnoisybandlimited} is based on high probability results for  the norm of a matrix restricted to random set of coordinates. For the purposes of such results, the uniform random sampling model and the Bernoulli sampling model where each component is taken independently and with equal probability is equivalent \cite{CandesRomberg_2006,CandesRombergTao_2006,TroppPaving_2008}. For instance, the derivation of  Theorem 1.2 of  \cite{candesRomberg_2007}, the main step of Lemma~\ref{lemUnitary:randomMeasFixedSupport}, is in fact based on a Bernoulli sampling model. Hence the high probability results presented in this lemma also hold for Gaussian erasure channel of Section~\ref{secUnitary:avg_erasureChannel} (with possibly different parameters).



\subsection{General Support} \label{secUnitary:probProjections}
In Section~\ref{secUnitary:probFixedDomain}, we have considered the case in which some of the eigenvalues of the covariance matrix are zero, and all the nonzero eigenvalues have the same value.  This case may be interpreted as the scenario where the signal to be estimated is exactly sparse. 
In this section, our aim is to find error bounds for estimation of not only sparse signals but also signals that are close to sparse.  Hence we are interested in the case where the signal has  small number of degrees of freedom effectively, that is when  a small portion of the eigenvalues carry most of the power of the signal. In this case,  the signal may not strictly have small number of degrees of freedom, but it can be well approximated by such a signal.  

We note that the result in this section makes use of a novel matrix theory result, and provides fundamental insights into problem of estimation of signals with small effective number of degrees of freedom. In the previous section we have used some results in compressive sensing literature that are directly applicable only when the signals have strictly small number of degrees of freedom (``insignificant'' eigenvalues of $K_x$ are exactly equal to zero.)  In this section we assume a more general eigenvalue distribution. Our result enables us draw conclusions when some of the eigenvalues are not exactly zero, but small.  The method of proof provides us a way to see the effects of the effective number of degrees of freedom of the signal ($\Lambda_x$) and the incoherence of measurement domain ($H U$), separately.

Before stating our result, we make some observations on the related results in random matrix theory.
Consider  the submatrices formed by restricting a matrix $K$ to random set of its rows, or columns;   $R_1 K$  or  $K R_2$ where $R_1$ and $R_2$ denote the restrictions to rows and columns respectively. The main tool for finding bounds on the eigenvalues of these submatrices is  finding a bound on  $\expectation ||R_1 K-\expectation[R_1 K] ||$  or  $\expectation ||K R_2^\dagger-\expectation[K R_2^\dagger] ||$\cite{Tropp_2008,Darses_2011,Tropp_2008short}. In our case such an approach is not very meaningful. The matrix we are  investigating $\Lambda_x^{-1}+(H U)^\dagger (H U)$ constitutes of two matrices: a deterministic diagonal matrix with possibly different entries on the diagonal and a random restriction.  Hence we adopt another method:  the approach of decomposing the unit sphere into compressible and incompressible vectors as proposed by M. Rudelson and R. Vershynin \cite{RudelsonVershynin_2008}.

We consider  the general measurement set-up in \eqref{eqnUnitary:linearSystem} where $ \orgum=  {{H}}  \unkum+ \nsum$, with  ${ K}_{\nsum} =\sigma_n^2 I_{M}$,  $K_x \succ 0$. The s.v.d. of $K_x$ is given as $K_x=U \Lambda_x U^\dagger$, where $U \in \mathbb{C}^{N \times N}$ is unitary and $\Lambda_x=\diag(\lambda_i)$ with $\sum_i \lambda_i=P$, $\lambda_1 \geq \lambda_2, \dots, \geq \lambda_N$.  $M$ components of $x$ are observed, where in each draw each component of the signal has equal probability of being selected. Hence the sampling matrix $H$ is a $M \times N$, $ M \leq N $ diagonal matrix, which may have repeated rows. This sampling scheme is slightly different than the sampling scheme of the previous section where the sampling locations are given by a set chosen uniformly at random from the set of all subsets of  $\{1,\ldots,N\}$  with size $M$. The differences in these models are very slight in practice, and we chose the former in this section due to the availability of partial uniform bounds on $||H U x||$ in this case. 


\begin{theorem}  \label{thm:hiiderror}
Let  $D(\delta)$ be the smallest number satisfying $\sum_{i=1}^D \lambda_i \geq \delta P$, where $\delta \in (0,1]$. Let ${\lambda_{max}}=\max_i{\lambda_i}=C_\lambda^{S}\, \frac{P}{D}$ and $\lambda_i <  C_\lambda^I\, \frac{P}{N-D}$, $i=D+1,\ldots,N$. Let $\mu(U) = \sqrt{N} \max_{k,j} |u_{k,j}| $. Let ${N}/{D}>\kappa \geq 1$. 
Let $\epsilon  \in (0,1)$,  $\theta \in (0, 0.5]$, and $\gamma \in (0,1)$. 
Let 
\begin{align}
M/\ln(10 M) \geq & C_1\, \theta^{-2} \mu^2 {\kappa D} \ln^2(100 {\kappa D}) \ln(4 N) \\
M \geq & C_2\  \theta^{-2} \mu^2 {\kappa D} \ln{(\epsilon^{-1})} \\
1 < & 0.5 \rho^2 \kappa \\
\rho \leq &  ({1-\gamma}) \frac{C_{\kappa D}}{C_{\kappa D}+1},
\end{align}
where 
\begin{equation}
C_{\kappa D} =  (1-\theta)^{0.5} \left(\frac{M}{N} \right)^{0.5}.
\end{equation}
Then the error will satisfy
\begin{align} \label{eqn:thm:Unitary}
& \prob  \biggl( \expectation [ ||x - \expectation [x | y ]||^2 ] \\
\nonumber
& \geq  (1-\delta) P +\max( \frac{P} {C_I} \, , \,  \frac{1}{\frac{1}{C_\lambda^{S}}+ \frac{1}{\sigma_n^2} \gamma^2 {C_{\kappa D}}^2 \frac{P}{ D} } P)  \biggr) \leq  \epsilon 
\end{align}
where 
\begin{equation}\label{eqn:CI}
C_I =  (0.5 \rho^2 \kappa -1) \,  \frac{0.5 \rho^2}{C_\lambda^{I} }   \frac{N-D}{N}. 
\end{equation}
Here $C_1 \leq 50\,963$ and $C_2 \leq 456$.
\end{theorem}

\begin{remark}
As we will see in the proof, the eigenvalue distribution plays a key role in obtaining stronger bounds: In particular, when the eigenvalue distribution is spread out, the theorem cannot provide bounds for low values of error. As the distribution becomes less spread out, stronger bounds are obtained.  We discuss these points after the proof the result. 
\end{remark}

{\bf{Proof:}} 
The error can be expressed as follows \eqref{eqUnitary:Woodburry}
\begin{align}
\nonumber
 & \expectation [ ||x- \expectation [x | y ]||^2 ] \\
 &=  \tr{((\Lambda_{x}^{-1}+\frac{1}{\sigma_n^2}  (HU) ^{\dagger} H U)^{-1})}   \\
  &= \sum_{i=1}^N \frac{1}{\lambda_i(\Lambda_x^{-1}  + \frac{1}{\sigma_n^2}(H U)^\dagger H U )} \\
 &= \sum_{i=1}^{N-D} \frac{1}{\lambda_i(\Lambda_x^{-1}  + \frac{1}{\sigma_n^2}(HU)^{\dagger}  H U )} \\
 \nonumber
 & \quad  + \sum_{i=N-D+1}^{N}  \frac{1}{\lambda_i(\Lambda_x^{-1}  + \frac{1}{\sigma_n^2} (H U)^\dagger H U )} \\
 \label{eqn:inveigupper}
  & \leq  \sum_{i=1}^{N-D} \frac{1}{\lambda_{i}(\Lambda_x^{-1} )} +\sum_{i=N-D+1}^{N}  \frac{1}{\lambda_i(\Lambda_x^{-1}  + \frac{1}{\sigma_n^2}(H U)^\dagger H U )} \\
  &\leq  \sum_{i=1}^{N-D} \lambda_{N-i+1}(\Lambda_x)+ D\frac{1}{\lambda_{min}(\Lambda_x^{-1}  + \frac{1}{\sigma_n^2}(H U)^\dagger H U )} \\
  &= \sum_{i=D+1}^{N} \lambda_{i}(\Lambda_x)+ D \frac{1}{\lambda_{min}(\Lambda_x^{-1}  + \frac{1}{\sigma_n^2}(H U)^\dagger H U )},   
 \end{align}
where \eqref{eqn:inveigupper} follows from case (a) of Lemma~\ref{lemUnitary:rankeigHorn}. 

Hence the error may be bounded as follows
 \begin{align} 
    \label{eqUnitary:ragleighErr}
 &  \expectation [ ||x - \expectation [x | y ]||^2 ]  \\
 \nonumber
  &  \leq  (1-\delta) P+ D \frac{1}{\lambda_{min}(\Lambda_x^{-1}  + \frac{1}{\sigma_n^2}(H U)^\dagger H U )} .
 \end{align}
The smallest eigenvalue of  $A = \Lambda_x^{-1}  + \frac{1}{\sigma_n^2}(HU) ^{\dagger}  H U$ is sufficiently away from zero with high probability as noted in the following lemma: 

\begin{lemma} \label{lemUnitary:rayleighEigenvalue} Under the conditions stated in Theorem  \ref{thm:hiiderror}, the eigenvalues of $A= \Lambda_x^{-1}  + \frac{1}{\sigma_n^2} (H U)^\dagger (H U) $ are bounded from below as follows:
\begin{align} \label{eqn:lemUnitary}
& \prob \bigl(\inf_{x\in S^{N-1}}  x^\dagger \Lambda_x^{-1} x + \frac{1}{\sigma_n^2} x^\dagger (H U) ^\dagger H U   x \\
\nonumber
& \leq \min ( {C_I} \, \frac{D}{P}, {\frac{1}{C_\lambda^{S} \frac{P}{D}}+ \frac{1}{\sigma_n^2} \gamma^2 {C_{\kappa D}}^2  } )\bigr) \leq \epsilon.
\end{align}
Here $S^{N-1}$ denotes the unit sphere where $x \in S^{N-1}$ if $x \in \mathbb{C}^N$, and $||x||=1$.
\end{lemma}

The proof of this lemma is given  in Section~\ref{proofUnitary:rayleighEigenvalue} of the Appendix. 

We now conclude the argument. 
Let us call the right-hand side of the eigenvalue bound in \eqref{eqn:lemUnitary} ${\bar{\lambda}}_{min}$. Then
\eqref{eqn:lemUnitary}  states that $\prob (\lambda_{min}(A) > {\bar{\lambda}}_{min})  \geq 1- \epsilon$, and hence we have the following: $\prob (\frac{1}{\lambda_{min}(A )} < \frac{1}{{\bar{\lambda}}_{min}})  \geq 1- \epsilon$.  Together with the error bound in \eqref{eqUnitary:ragleighErr}, we have $ \prob ( \expectation [ ||x - \expectation [x | y ]||^2 ]  <  (1-\delta) P+ D \frac{1}{\bar{\lambda}}_{min}) \geq 1- \epsilon$, and the result follows. \myqed

%
We now discuss the error bound that Theorem~\ref{thm:hiiderror} provides. The expression in \eqref{eqn:thm:Unitary} can be interpreted as an upper bound on the error that holds with probability at least $1-\epsilon$. The bound consists of a $(1-\delta) P$ term and a $\max$ term. This $(1-\delta) P$  term is the total power in the eigenvalues that are considered to be insignificant (i.e. $\lambda_i$ such that $i \notin \mathcal{D}=\{1,\ldots,D\}$).  This term is a bound for  the error that would have been introduced if we had preferred not estimating the random variables corresponding to these insignificant eigenvalues. 
Since in our setting we are interested in signals with effectively small number of degrees of freedom, hence $\delta$ close 
 to $1$  for $D$ much smaller than $N,$  this term will be typically small.
Let us now look at the term that will come out of the maximum function. 
When the noise level is relatively low, the $\frac{P} {C_I}$ term comes out of the $\max$ term. 
Together with the $\rho$ and $\kappa$ whose choices will depend on $D$,  order of magnitude of this term substantially depends on the value of the insignificant eigenvalues. This term  may be interpreted as an upper bound on the error due to the random variables associated with the insignificant eigenvalues acting as noise for estimating of the random variables associated with the significant eigenvalues (i.e.  $\lambda_i$ such that $i \in \mathcal{D}$). Hence in the case where the noise level is relatively low, the random variables associated with the insignificant eigenvalues become the dominant source of error in estimation.
By choosing $\kappa$ and $\gamma$ appropriately, this term can be made small provided that $D$ is small compared to $N$, which is the typical scenario we are interested in.  
When the noise level is relatively high,  the second argument comes out of the $\max$ term. Hence for relatively high levels of noise, system noise $n$ rather than the signal components associated with the insignificant eigenvalues becomes the dominant source of error in the estimation. This term can  be also written as 
\begin{align}
 \frac{1}{\frac{1}{C_\lambda^{S}}+ \frac{1}{\sigma_n^2} \gamma^2 {C_{\kappa D}}^2 \frac{P}{ D} } P= &
  \frac{1}{\frac{1}{C_\lambda^{S}}+ \frac{1}{\sigma_n^2} \gamma^2 (1-\theta) \frac{M}{N} \frac{P}{ D} } P \\
    = & \frac{1}{\frac{1}{C_\lambda^{S}}+  \gamma^2 (1-\theta)\, \text{SNR} } P,
    \end{align}
where $\text{SNR}= \frac{1}{{\sigma^2_n}} \frac{P}{D}  \frac{M}{N}$. We note that the general form of this expression is the same as the general form of the error expression in Section~\ref{secUnitary:probFixedDomain} (see \eqref{eqnUnitary:randomMeasFixedSupport:disc}), where the error bound is of the  general form $\frac{1}{1+ c SNR} P$, where $c \in (0,1]$. In  Section~\ref{secUnitary:probFixedDomain}, the case where the signal have exactly small number of degrees of freedom with $D$ is considered, in which case $C_\lambda^{S} = 1$, $\delta=1$ and $D=|B|$.  
We observe that here,  there are two factors that forms the effective SNR loss $c=\gamma^2 (1-\theta)$. A look through the proof (in particular, Lemma~\ref{lem:Bauhutsparseunitary} and Lemma~\ref{lem:compunitary}) reveals that  the effective SNR loss due to $(1-\theta) $ factor is the term that would have been introduced if we were to work with signals where $ \kappa D$ eigenvalues are equal and nonzero, and the others zero.
This factor also introduces a loss of SNR due to considering signals with $\kappa D, \kappa>1$ instead $D$ nonzero eigenvalues. The $\gamma^2$ term may be interpreted as an additional loss due to working with signals for which  $\lambda_i$ such that $i \notin \mathcal{D}$ are not zero.




\section{Conclusions}\label{secUnitary:conc}
We have considered the transmission of a Gaussian vector source over a multi-dimensional Gaussian channel where a random or a fixed subset of the channel outputs are erased.  The unitary transformation that connects the canonical signal domain and the measurement space played a crucial role in our investigation. Under the assumption the estimator knows the channel realization, we have investigated the MMSE performance, both in average, and also in terms of guarantees that hold with high probability as a function of system parameters.

We have considered the sampling model of random erasures. 
We have considered two channel structures:  i) random Gaussian scalar channel where only one measurement is done through Gaussian noise  and ii) vector channel where measurements are done through parallel Gaussian channels with a given channel erasure probability. Under these channel structures, we have formulated the problem of finding the most favorable unitary transform under average (w.r.t. random erasures) MMSE criterion. We have  investigated the convexity properties of this optimization problem, and obtained necessary conditions of optimality through variational equalities. We were not able to solve this problem in its full setting, but we have solved some related special cases. Among these we have identified special cases where DFT-like unitary transforms (unitary transforms with  $|u_{ij}|^2=\frac{1}{N}$) turn out to be the best coordinate transforms, possibly along with other unitary transforms. Although these observations and the observations of Section~\ref{secUnitary:probFixedDomain} (which are based on compressive sensing results) may suggest that the DFT is optimal in general, we showed through a counterexample that this is not the case under the performance criterion of average MMSE.

In Section \ref{secUnitary:prob}, we have focused on  performance guarantees that hold with high probability.  We have presented upper bounds on the MMSE depending on the support size and the number of measurements. We have also considered more general eigenvalue distributions, (i.e. signals that may not strictly have low degree of freedom, but effectively do so), and we have illustrated the interplay between the amount of information in the signal, and the spread of this information in the measurement domain for providing performance guarantees.

To serve as a benchmark, we have considered sampling of circularly wide-sense stationary signals, which is a natural way to model wide-sense stationary signals in finite dimension. Here the covariance matrix was circulant by assumption, hence the unitary transform was fixed and given by the DFT matrix. 
We have focused on the commonly employed equidistant sampling strategy and gave the explicit expression for the MMSE.

In addition to providing insights into the problem of unitary encoding in Gaussian erasure channels, our work in this article also contributed to our understanding of the relationship between the MMSE and  the total uncertainty in the signal as quantified by  information theoretic measures such as entropy  (eigenvalues) and the spread of this uncertainty (basis). 
We believe that through this relationship our work also sheds light on how to properly characterize the concept of  ``coherence of a random field".  Coherence, a concept describing the overall correlatedness of a random field, is of central importance in statistical optics; see for example \cite{b_mandelWolf,ozaktas_2002} and the references therein.

\section*{Acknowledgement}
The authors thank the Associate Editor and the anonymous reviewers for their helpful comments. In particular, we thank the Associate Editor for pointing out a shorter proof for minimizing the expression given in \eqref{eqn:zk}.

\appendices

\section{Notes on equidistant sampling of c.w.s.s. signals}\label{sec:pf:lemUnitary:mmsecwss}
We believe that error expressions related to the equidistant sampling of the c.w.s.s. signals can be also of independent interest. Hence we further elaborate on this sampling scenario in this section. 
We first present the result for the noiseless case and then give the relevant proofs, including that of  Lemma {\protect \ref{lemUnitary:rayleighEigenvalue}} which is for the noisy sampling case.

\subsection{Equidistant sampling without noise}
 Our set-up is the same with Section~\ref{secUnitary:cwss} except here we first consider the case where there is no noise so that $y=H x$. 
We now present an explicit expression and an upper bound for the mean-square error  associated with this noiseless  set-up. 
\begin{lemma} \label{lemUnitary:mmse_cwss} Let the model and  the sampling strategy be as described above.
Then the MMSE of estimating $x$ from these equidistant samples can be expressed as
\begin{align} \label{eqnUnitary:mmse_cwssmain}
 & \expectation [ ||x - \expectation [x | y ]||^2 ] \\
 \nonumber
 &=   \sum_{k \in J_0} ( \sum_{i=0}^{\Delta N-1} \lambda_{i M +k} -\sum_{i=0}^{\Delta N-1}  \frac{\lambda_{i M +k}^2}{\sum_{l=0}^{\Delta N-1} \lambda_{l M +k}}),
 \end{align} 
where   $J_0 =\{k: {\sum_{l=0}^{\Delta N-1} \lambda_{l M +k}} \neq  0, \,\, 0 \leq k  \leq M-1 \}  \subseteq \{0, \ldots, M-1 \}$.

In particular, choose a set of indices $J \subseteq \{0,1,\dots, N-1\} $ with  $|J|=M$ such that $\forall i,j ,\,\,\, 0 \leq i, j \leq \Delta N-1 , i \neq j$
 \begin{equation}\label{eqnUnitary:setJ}
 {jM+k}  \in J  \Rightarrow  {iM+k} \notin J  \quad  
\end{equation}
with $0 \leq k \leq M-1$. Let $ P_J=\sum_{i \in J} \lambda_i$.
Then the  MMSE   is upper bounded by the total power in the remaining eigenvalues
\begin{align}
 \expectation [ ||x- \expectation [x | y ]||^2 ]   \leq  2 (P -P_J).
\end{align}
In particular, if there is such a set $J$ so that $P_J=P$, the MMSE will be zero.
\end{lemma}

\begin{remark}
The set $J$ essentially consists of the indices which do not overlap when shifted by $M$.
\end{remark}

\begin{remark}
We note that the choice of the set $J$ is not unique, and each choice of the set of indices may provide a different upper bound. To obtain the lowest possible upper bound,  one should consider the set with the largest total power.
\end{remark}

\begin{remark}
If there exists such a set $J$ that has the most of power, i.e.
$P_J  = \delta P$,  $\delta \in (0,1]$,  with $\delta$ close to 1, then $2 (P -P_J)=2 (1-\delta)P $ is small and the signal can be estimated with low values of error.
In particular, if such a set has all the power, i.e. $P=P_J$, the error will be zero.
A conventional aliasing free set  $J$ may be the set of indices of the band of a  band-pass signal with a band smaller than $M$.  It is important to note that there may exist other sets $J$ with $P=P_J$, hence the signal may be aliasing free even if the signal is not bandlimited (low-pass, high-pass etc) in the conventional sense.
\end{remark}

{\bf{Proof:}} Proof is given in Section~\ref{proofUnitary:mmse_cwss} of the Appendix.

We observe that the bandwidth  (or the effective degrees of freedom) turn out to be good predictors of estimation error in equidistant sampling scenario. On the other hand, the differential entropy of an effectively bandlimited Gaussian vector can be very small even if the bandwidth is close to $N$, hence may not provide any useful information with regards to estimation performance.

We now compare our error bound with the related results in the literature. In the following works, similar problems with signals defined on $\mathbb{R}$ are considered:  In \cite{Brown_1978}, mean-square error of approximating  a possibly non-bandlimited wide-sense stationary (w.s.s.) signal using sampling expansion is considered and  a uniform upper bound  in terms  of power outside the bandwidth of approximation is derived. Here we are interested in the average error over all points of the $N$ dimensional vector.  Our method of approximation of the signal is possibly different, since we use the MMSE estimator. As a result our bound also makes use of the shape of the eigenvalue distribution. \cite{Lloyd_1959} states that a w.s.s. signal is determined linearly by its samples if some set of frequencies containing all of the power of the process is disjoint from each of its translates where the amount of translate is determined by the sampling rate. Here for circularly w.s.s. signals we show a similar result: if there is a set $J$ that consists of indices which do not overlap when shifted by $M$, and has all the power, the error will be zero. In fact, we show a more general result for our set-up and give the explicit error expression. We also show that two times the power outside this set $J$ provides an upper bound for the error, hence putting a bound on error even if it is not exactly zero.

\subsection{Proof of Lemma~{\protect \ref{lemUnitary:mmse_cwss}}} \label{proofUnitary:mmse_cwss}
We remind that in this section  $u_{tk}={1 \over \sqrt N} e^{j  {2 \pi \over N} t k}$, $0 \leq t \, , k \leq N-1 $ and the associated eigenvalues are denoted with $\lambda_k$ without  reindexing them  in decreasing/increasing order. We first assume that $K_y= \expectation{[y y^\dagger]} =H K_x H^\dagger$  is non-singular. The generalization to the case where $K_y$ may be singular is presented at the end of the proof.

The MMSE for estimating $x$ from $y$ is given by \cite[Ch.2]{b_andersonMoore_optFiltering}
\begin{dgroup*}
\begin{dmath*}
 \expectation[ ||x - \expectation [x | y ] ||^2 ] = \tr(K_x-K_{xy} K_y^{-1} K_{xy} ^\dagger) 
 \end{dmath*}
   \begin{dmath} \label{eqnUnitary:circErr}
     = \tr(\Lambda _x-\Lambda_x U^\dagger H^\dagger  (H U \Lambda_x U^\dagger H^\dagger)^{-1} H U \Lambda_x ).
     \end{dmath}
\end{dgroup*}
We now consider $H U \in \mathbb{C}^{M \times N}$, 
\begin{eqnarray}
(H U)_{lk} = {1 \over \sqrt N} e^{j  {2 \pi \over N} (\Delta N l)  k} = {1 \over \sqrt N} e^{j  {2 \pi \over {M}}  l k},
 \end{eqnarray} 
where $0 \leq l \leq {N \over \Delta N}-1$,\,\,\, $0 \leq k \leq N-1$. We  observe that for a given  $l$, $e^{j  {2 \pi \over {M}} l k}$ is a periodic function of $k$ with period $M={N \over \Delta N}$. Hence, $l^{th}$ row of $H U$ can be expressed as
\begin{align*}
(H U)_{l:} &=  \frac{1}{\sqrt{N}}\, [e^{j  {2 \pi \over {{M}}}  l [0 \ldots {N}-1]}]\\ 
&=  \frac{1}{\sqrt{N}}\, [e^{j  {2 \pi \over {M}}  l [0 \ldots {M}-1]}| \ldots| e^{j  {2 \pi \over {M}} l [0 \dots {M}-1]}].
\end{align*}
Let $U_M$ denote the $M \times M$ DFT matrix, i.e. $(U_M)_{l k}={1 \over \sqrt{M}} e^{j  {2 \pi \over {M}} l k}$ with  $0 \leq l \leq M-1$,\,\,\, $0 \leq k \leq M-1$. Hence $H U $ is the matrix formed by stacking $\Delta N$  $M \times M$ DFT matrices side by side
\begin{align}
H U= \frac{1} {\sqrt {\Delta N}} [ U_M |\ldots| U_M ].
 \end{align} 

Now we consider the covariance matrix of the observations $K_y= H K_x H^\dagger=H U \Lambda_x U^\dagger H^\dagger$. We first express $\Lambda_x$ as a block diagonal matrix as follows 
\begin{align*}
\Lambda_x &=\left[\begin{array}{cccc}\lambda_0 & 0 & \cdots & 0 \\0 & \lambda_1 &  & \vdots \\  \vdots &   & \ddots & \vdots \\0 & \cdots &  0 & \lambda_{N-1} \end{array}\right] \\
&= \left[\begin{array}{cccc}\Lambda_x^0 & {\bar 0} & \cdots & {\bar 0} \\{\bar 0} & \Lambda_x^1 &  & \vdots \\  \vdots &  & \ddots & \vdots \\{\bar 0} & \cdots &  {\bar 0} & \Lambda_x^{\Delta N -1}\end{array}\right].
\end{align*}
where ${\bar 0} \in \mathbb{R}^{M \times M}$ denotes the matrix of zeros. 
Hence 
$\Lambda_x=\diag(\Lambda_x^i)$ with $\Lambda_x^i=\diag(\lambda_{i M+k}) \in \mathbb{R}^{M \times M}$, where $ 0 \leq i \leq \Delta N-1$,\,\, $0 \leq k \leq M-1$. We can write $K_y$ as
\begin{align*}
K_y &= H U \Lambda_x U^\dagger H^\dagger\\
      &= \frac{1} {\sqrt {\Delta N}} [ U_M |\ldots| U_M ]  \diag(\Lambda_x^i) \left[\begin{array}{c}U_M^\dagger  \\ \vdots \\ U_M^\dagger \end{array}\right] \frac{1} {\sqrt {\Delta N}} \\
       &=\frac{1} { {\Delta N}}  \,\, U_M ( \sum_{i=0}^{\Delta N-1} \Lambda_x^i) U_M^\dagger
 \end{align*} 
 We note that $ \sum_{i=0}^{\Delta N-1} \Lambda_x^i \in \mathbb{R}^{M \times M}$ is formed by summing diagonal matrices, hence also diagonal. Since $U_M$ is the $M \times M$ DFT matrix,  $K_y$ is again a circulant matrix whose $k^{th}$ eigenvalue is given by 
 \begin{align} \label{eqnUnitary:Kyeigenvalues}
 \lambda_{y,k}= \frac{1} { {\Delta N}}   \sum_{i=0}^{\Delta N-1} \lambda_{i M+k}, \quad 0 \leq k \leq M-1.
 \end{align}
Hence $K_y=  U_M \Lambda_y U_M^\dagger $ is the eigenvalue-eigenvector decomposition of $K_y$, where  $\Lambda_Y =\frac{1} { {\Delta N}}   \sum_{i=0}^{\Delta N-1} \Lambda_x^i=\diag(\lambda_{y,k})$. 
There may be aliasing in the eigenvalue spectrum of $K_y$ depending on the eigenvalue spectrum of $K_x$ and $\Delta N$. We also note that $K_y$ may be aliasing free even if it is not bandlimited (low-pass, high-pass, etc.) in the conventional sense. We note that since $K_y$ is assumed to be non-singular, $\lambda_{y,k}>0$. 
$K_y^{-1}$ can be expressed as
\begin{align*}
K_y^{-1} &=  (U_M  \Lambda_y U_M^\dagger)^{-1}\\
              &=  U_M  \diag(\frac{1}{\lambda_{y,k}}) U_M^\dagger\\
             &=  U_M  \diag({\Delta N \over{  \sum_{i=0}^{\Delta N-1} \lambda_{i M+k}}}) U_M^\dagger.
 \end{align*} 
 We are now ready to consider the error expression in \eqref{eqnUnitary:circErr}. We first consider the second term, that is
\begin{align*}
\tr (\Lambda_x U^\dagger & H^\dagger  K_y^{-1} H U \Lambda_x )\\
& =  \tr( \frac{1}{\sqrt {\Delta N}}  \left[\begin{array}{c}\Lambda_x^0 U_M^\dagger  \\ \vdots \\ \Lambda_x^{\Delta N-1}U_M^\dagger \end{array}\right]   ( U_M \Lambda_y^{-1} U_M^\dagger) \\
& \times \frac{1}{\sqrt {\Delta N}} [ U_M \Lambda_x^0 |\ldots| U_M \Lambda_x^{\Delta N-1} ] )  \\
 &= \sum_{i=0}^{\Delta N-1}  \frac{1}{\Delta N}  \tr (\Lambda_x^i \Lambda_y^{-1} \Lambda_x^i) \\
  &= \sum_{i=0}^{\Delta N-1} \sum_{k=0}^{M-1} \frac{\lambda_{i M +k}^2}{\sum_{l=0}^{\Delta N-1} \lambda_{l M +k}}
 \end{align*}

Hence the MMSE becomes 
\begin{align*}
 & \expectation [ ||x - \expectation [x | y ]||^2 ]  \\
 &= \sum_{t=0}^{N-1} \lambda_t  - \sum_{i=0}^{\Delta N-1} \sum_{k=0}^{M-1} \frac{\lambda_{i M +k}^2}{\sum_{l=0}^{\Delta N-1} \lambda_{l M +k}}\\
 &= \sum_{k=0}^{M-1}  \sum_{i=0}^{\Delta N-1} \lambda_{i M +k}  - \sum_{i=0}^{\Delta N-1} \sum_{k=0}^{M-1} \frac{\lambda_{i M +k}^2}{\sum_{l=0}^{\Delta N-1} \lambda_{l M +k}}\\
 &=  \sum_{k=0}^{M-1} ( \sum_{i=0}^{\Delta N-1} \lambda_{i M +k} -\sum_{i=0}^{\Delta N-1}  \frac{\lambda_{i M +k}^2}{\sum_{l=0}^{\Delta N-1} \lambda_{l M +k}}).
 \end{align*} 
We note that we have now expressed the MMSE as the sum of the errors in $M$ frequency bands.
Let us define the error at $k^{th}$ frequency band as 
\begin{align}\label{eqn:ekw}
e_k^w = \sum_{i=0}^{\Delta N-1} \lambda_{i M +k} -\sum_{i=0}^{\Delta N-1}  \frac{\lambda_{i M +k}^2}{\sum_{l=0}^{\Delta N-1} \lambda_{l M +k}},
\end{align}
where  $0 \leq k \leq M-1$. Hence the total error is given by
\begin{align*}
 \expectation [ ||x - \expectation [x | y ]||^2 ]   = \sum_{k=0}^{M-1} e_k^w.
 \end{align*}
 That proves the expression for the error.  We now consider the upper bound. Before moving on, we study a special case:

{\ex
 \label{ex:DeltaN2} Let $\Delta N=2$.  Then 
\begin{align*}
e_k^w &= \lambda_k+\lambda_{{N \over 2} +k} -\frac{\lambda_k^2+\lambda_{{N \over 2} +k}^2}{\lambda_k+\lambda_{{N \over 2} +k} } \\
         &= \frac{2 \lambda_k \lambda_{{N \over 2} +k}}{\lambda_k+\lambda_{{N \over 2} +k} }.
 \end{align*}
Hence $   {1 \over e_k^w} =   {1 \over 2} ( {1 \over \lambda_{{N \over 2} +k} } + {1 \over \lambda_{k}})$.
We note that this is the MMSE for the following single output multiple input system 
\begin{align}\label{eqnUnitary:2by2channel}
z^k =
\left[\begin{array}{cc} {1} & 1  \end{array}\right]\,  \left[\begin{array}{c} s_0^k \\ s_1^k  \end{array}\right], 
\end{align}
where $s^k~\sim\mathcal{N}(0,K_{s^k})$, with $K_{s^k}=\diag (\lambda_k,\lambda_{{N \over 2} +k} )$. Hence the random variables associated with the frequency components at  $k$, and ${N \over 2} +k$ act as interference for estimating the other one. We observe that for estimating $x$ we have $N \over 2$ such channels in parallel.

We may bound $e_k^w$ as       
  \begin{align*}        
e_k^w  = \frac{2 \lambda_k \lambda_{{N \over 2} +k}}{\lambda_k+\lambda_{{N \over 2} +k} } 
      &\leq  \frac{2 \lambda_k \lambda_{{N \over 2} +k}}{\max (\lambda_k,\lambda_{{N \over 2} +k} )} \\
         &= 2 \min (\lambda_k,\lambda_{{N \over 2} +k} ).
\end{align*}
This bound may be interpreted as follows: Through the scalar channel shown in \eqref{eqnUnitary:2by2channel}, we would like to learn two random variables $s_0^k$  and $s_1^k$. The error of this channel is upper bounded by the error of the scheme where we only  estimate the one with the largest variance, and don't try to estimate the variable with the small variance. In that scheme, one first makes an error of  $\min (\lambda_k,\lambda_{{N \over 2} +k})$,  since the variable with the small variance is ignored.  We may lose another   $ \min (\lambda_k,\lambda_{{N \over 2} +k}) $, since this variable acts as additive noise for estimating the variable with the larger variance, and the MMSE associated with such a channel may be upper bounded by the variance of the noise.

Now we choose the set of indices $J$ with $|J|=N/2$ such that  $k \in J  \Leftrightarrow  {{N \over 2} +k} \notin J$   and $J$ has the most power over all such sets, i.e.  $\displaystyle k+ \arg \max_{k_0 \in \{0,N/2\}} \lambda_{k_0+k} \in J$, where $0 \leq k \leq N/2-1$. Let $ \displaystyle P_J = \sum_{k \in J} \lambda_k$. 
Hence
\begin{align*}
 \expectation [ ||x - \expectation [x | y ]||^2 ]   =  \sum_{k=0}^{N/2-1}  e_k^w 
  & \leq    2 \sum_{k=0}^{N/2-1}  \min (\lambda_k,\lambda_{{N \over 2} +k} )  \\
   & = 2 (P-P_J).
\end{align*} 
We observe that the error is upper bounded by $2 \times $ (the power in the ``ignored band"). 
}

\vspace{10pt}
We now return to the general case. 
Although it is possible to consider any set $J$ that satisfies the assumptions stated in \eqref{eqnUnitary:setJ}, for notational convenience we choose the set $J=\{0,\ldots,M-1\}$. Of course in general one would look for the set $J$ that has most of the power in order to have a stricter bound on the error. 

We consider  \eqref{eqn:ekw}. We note that this is the MMSE of estimating $s^k$  from the output of the following single output multiple input  system
\begin{align*}
z^k =
\left[\begin{array}{ccc} {1} & \cdots &1  \end{array}\right]\,  \left[\begin{array}{c} s_{1}^k \\ \vdots  \\ {s_{\Delta N-1}^k} \\  \end{array}\right],
\end{align*}
where  $s^k~\sim\mathcal{N}(0,K_{s^k})$, with $K_{s^k}$ as follows
\begin{align*}
K_{s^k} &=\diag(\sigma_{s_{i}^k}^2) \\
&=\diag (\lambda_{k},\dots, \lambda_{i M +k} , \dots ,\lambda_{({{\Delta N-1}}) M +k} ).
\end{align*}
We define 
\begin{align*} 
P^k={\sum_{l=0}^{\Delta N-1} \lambda_{l M +k}},\quad \quad    0 \leq k \leq M-1
\end{align*}
We note that $ \sum_{k=0}^{M-1}  P^k=P$.

We now bound $e_k^w$ as in the  $\Delta N=2$ example
\begin{align*} 
e_k^w &= \sum_{i=0}^{\Delta N-1} \lambda_{i M +k} -\sum_{i=0}^{\Delta N-1}  \frac{\lambda_{i M +k}^2}{\sum_{l=0}^{\Delta N-1} \lambda_{l M +k}}  \\
         &= \sum_{i=0}^{\Delta N-1} (\lambda_{i M +k} - \frac{\lambda_{i M +k}^2}{P^k})\\
           &= (\lambda_{k} - \frac{\lambda_{k}^2}{P^k}) + \sum_{i=1}^{\Delta N-1} (\lambda_{i M +k} - \frac{\lambda_{i M +k}^2}{P^k})\\ 
           &\leq  ( P^k -\lambda_k)+  \sum_{i=1}^{\Delta N-1}  \lambda_{i M +k} \\
            &=  ( P^k -\lambda_k)+  P^k- \lambda_k \\
            &=  2( P^k -\lambda_k),
\end{align*}
where we have used $\lambda_{k} - \frac{\lambda_{k}^2}{P^k} = \frac{\lambda_{k} ( P^k-\lambda_{k})}{P^k}
 \leq   P^k-\lambda_{k}$ 
since $ 0 \leq \frac{\lambda_k}{P^k} \leq 1$
and $\lambda_{i M +k} - \frac{\lambda_{i M +k}^2}{P^k} \leq \lambda_{i M +k} $ 
since $ \frac{\lambda_{i M +k}^2}{P^k} \geq 0$. This upper bound may interpreted similar to the Example \ref{ex:DeltaN2}: The error is upper bounded by the error of the scheme where one estimates the random variable associated with $\lambda_k$, and ignore the others.   

The total error is bounded by 
\begin{align*}
 \expectation [ ||x - \expectation [x | y ]||^2 ]   = \sum_{k=0}^{M-1} e_k^w 
	& \leq  \sum_{k=0}^{M-1} 2( P^k -\lambda_k) \\
	& =  2 (\sum_{k=0}^{M-1}  P^k - \sum_{k=0}^{M-1}  \lambda_k) \\
	& =  2 (P - P_J).
 \end{align*} 

\begin{remark}
We now consider the case where $K_y$ may be singular. In this case, for MMSE estimation, it is enough to use $K_y^{+}$ instead of $K_y^{-1}$, where $^{+}$ denotes the Moore-Penrose pseudo-inverse \cite[Ch.2]{b_andersonMoore_optFiltering}. Hence the MMSE may be expressed as  $\tr(K_x-K_{xy} K_y^{+} K_{xy} ^\dagger)$. We have $K_y^{+} = (U_M  \Lambda_y U_M^\dagger)^{+}=U_M  \Lambda_y^{+} U_M^\dagger = U_M  \diag({\lambda_{y,k}}^{+}) U_M^\dagger$, where $\lambda_{y,k}^{+}=0$ if $\lambda_{y,k}=0$ and $\lambda_{y,k}^{+}=\frac{1}{\lambda_{y,k}}$ otherwise. Going through the calculations with $K_y^{+}$ instead of $K_y^{-1}$ reveals that the error expression remains essentially the same
\begin{align*}
 & \expectation [ ||x - \expectation [x | y ]||^2 ] \\
 &=  \sum_{k \in J_0} ( \sum_{i=0}^{\Delta N-1} \lambda_{i M +k} -\sum_{i=0}^{\Delta N-1}  \frac{\lambda_{i M +k}^2}{\sum_{l=0}^{\Delta N-1} \lambda_{l M +k}}),
 \end{align*} 
where   $J_0 =\{k: {\sum_{l=0}^{\Delta N-1} \lambda_{l M +k}} \neq     0, 0 \leq k  \leq M-1 \}  \subseteq \{0, \ldots, M-1 \}$. We note that $\Delta N \lambda_{y,k} ={\sum_{l=0}^{\Delta N-1} \lambda_{l M +k}}=P^k$.  
\end{remark}

\subsection{Proof of Lemma~\protect{\ref{lemUnitary:mmse_cwssnoisy}}}
The proof of Lemma~\ref{lemUnitary:mmse_cwssnoisy} follows from the proof of Lemma~\ref{lemUnitary:mmse_cwss}  as follows: 
We first  note that in the noisy case $K_{xy}= K_x H^\dagger$, as in the noiseless case.
We also note that in the noisy case, $K_y$ is given by $K_y= H K_x H^\dagger + K_n$. 
Now the result is obtained by retracing the steps of the proof of  Lemma~\ref{lemUnitary:mmse_cwss}, which is given in Section~\ref{proofUnitary:mmse_cwss},  with $K_y$ replaced by the above expression, that is $K_y= H K_x H^\dagger + K_n$.


\section{Proof of Lemma {\protect \ref{lemUnitary:rayleighEigenvalue}}} \label{proofUnitary:rayleighEigenvalue} 
Our aim is to show that the smallest eigenvalue of $A=\Lambda_x^{-1}+\frac{1}{\sigma_n^2} (HU) ^{\dagger}  H U$ is   bounded from below with a sufficiently large number with high probability. That is, we are interested in  
\begin{align} \label{eqnUnitary:inf}
\inf_{x\in S^{N-1}}  x^\dagger \Lambda_x^{-1} x + \frac{1}{\sigma_n^2} x^\dagger (HU) ^{\dagger}  H U x.
\end{align} 
To lower bound the smallest eigenvalue, we adopt the approach proposed by  \cite{RudelsonVershynin_2008}: We consider the decomposition of the unit sphere into two sets,  compressible vectors and incompressible vectors.   We recall the following from \cite{RudelsonVershynin_2008}:

\begin{definition} [pg.14, \cite{RudelsonVershynin_2008}] Let  $|\text{supp}(x)|$ denote the number of elements in the support of $x$.  Let $\eta, \rho \in (0,1)$. $x \in \mathbb{C^N}$ is sparse,  if $|\text{supp}(x)| \leq \eta N$. The set of vectors sparse with a given $\eta$   is denoted by $Sparse (\eta)$.
 $x  \in S^{N-1}$ is compressible,  if $x$ is within an Euclidean distance $\rho$ from the set of all sparse vectors, that is $\exists\, y \in \text{Sparse} (\eta), d(x,y) \leq \rho$. The set of compressible vectors is denoted by   $Comp(\eta,\rho)$.
 $x \in S^{N-1}$ is incompressible if it is not compressible. The set of incompressible vectors is denoted by $Incomp(\eta,\rho)$.
 \end{definition} 
 
 \begin{lemma} \label{IncompSpread} [Lemma 3.4, \cite{RudelsonVershynin_2008}] Let $x \in Incomp(\eta,\rho)$. Then there exists a set  $\psi \subseteq \{1,...,N\}$ of cardinality $|\psi| \geq 0.5 \rho^2 \eta N$ such that  
 \begin{eqnarray}
 \frac{\rho}{\sqrt{2 N}} \leq |x_k| \leq \frac{1}{\sqrt{\eta N}}, \quad \quad \forall k \in \psi.
 \end{eqnarray}
 \end{lemma}

The set of compressible and incompressible vectors provide a decomposition of the unit sphere, i.e.  $S^{N-1}= Incomp(\eta,\rho) \bigcup  Comp(\eta,\rho)$ \cite{RudelsonVershynin_2008}.   We will show that the first/second term in $\eqref{eqnUnitary:inf}$ is sufficiently away from zero for $x \in Incomp(\eta,\rho)$/ $x \in Comp(\eta,\rho)$ respectively. The parameters $\rho$ and $\eta  = \kappa D/N$, $\kappa>1$ are going to be chosen appropriately to satisfy the conditions of Lemma {\ref{lemUnitary:rayleighEigenvalue}}.

As noted in \cite{RudelsonVershynin_2008}, for any square matrix $A$
\begin{align}
\nonumber
\prob (\inf_{x\in S^{N-1}}  x^\dagger A x \leq C ) 
& \leq \prob (\inf_{x \in Comp(\eta,\rho)} x^\dagger A x \leq C )\\
  \label{eqUnitary:decomp}
& + \prob (\inf_{x \in Incomp(\eta,\rho) }  x^\dagger A x \leq C).
\end{align} 
We also note that 
\begin{align}
\nonumber
 & \inf_{x \in Incomp(\eta,\rho)}   x^\dagger \Lambda_x^{-1} x + x^\dagger \frac{1}{\sigma_n^2} (HU) ^{\dagger}  H U x  \\
\nonumber
 & \geq \inf_{x \in Incomp(\eta,\rho)}  x^\dagger \Lambda_x^{-1} x   \\
\label{eqUnitary:decompincomp}
&= \inf_{x \in Incomp(\eta,\rho)}  ||\Lambda_x^{-1/2} x ||^{2},
\end{align}
and
\begin{align}
\nonumber
& \inf_{x \in Comp(\eta,\rho)}  x^\dagger \Lambda_x^{-1} x + x^\dagger \frac{1}{\sigma_n^2} (HU) ^{\dagger}  H U x \\
\nonumber
& \geq \frac{1}{\lambda_{max}}+\inf_{x \in Comp(\eta,\rho)} x^\dagger \frac{1}{\sigma_n^2} (HU) ^{\dagger}  H U x  \\
 \label{eqUnitary:decompcomp}
&=  \frac{1}{\lambda_{max}} + \frac{1}{\sigma_n^2} (\inf_{x \in Comp(\eta,\rho)} ||H U  x||^2 ),
\end{align}
where  ${\lambda_{max}}=\max_i{\lambda_i}$ and the inequalites are due to the fact that $ \Lambda_x^{-1}$, $H ^\dagger H$ are both positive-semidefinite.

We now recall the following result from \cite{Rauhut_2010}, which expresses the eigenvalue bound for sparse vectors. 
\begin{lemma}\label{lem:Bauhutsparseunitary}\cite[Theorem 8.4]{Rauhut_2010}
Let $U$ be an $N \times N$ unitary matrix with  $\mu = \sqrt{N} \max_{k,j} |u_{k,j}| $.
Let $\epsilon  \in (0,1)$,  ${\theta}_{\eta} \in (0, 0.5]$. 
If  
\begin{align}
M/\ln(10 M) \geq & C_1\, { \theta}_{\eta}^{-2} \mu^2 {\kappa D} \ln^2(100 {\kappa D}) \ln(4 N) \\
M \geq & C_2\  { \theta}_{\eta}^{-2} \mu^2 {\kappa D} \ln{\epsilon^{-1}}
\end{align}
Then,  
\begin{equation}
P (\inf_{x \in Sparse(\eta)} ||H U x||^2  \leq ({1-{ \theta}_{\eta}}) \frac{M}{N} ||x||^2 ) \leq \epsilon.
\end{equation}
 Here $C_1 \leq 50\,963$, $C_2 \leq 456$ and $\eta = \kappa D/N$.
\end{lemma} 

We now show that this result can be generalized to an eigenvalue bound for compressible vectors $x \in Comp(\eta,\rho)$, where $\rho$ will be appropriately chosen.  
\begin{lemma}\label{lem:compunitary}
Let the conditions of Lemma~\ref{lem:Bauhutsparseunitary} hold. 
Let $C_{\kappa D} = ({1-{ \theta}_{\eta}})^{0.5} (\frac{M}{N})^{0.5} $. Choose $\rho$ such that 
\begin{equation}
\rho \leq  ({1-\gamma}) \frac{C_{\kappa D}}{C_{\kappa D}+1},
\end{equation}
where $0 \leq \gamma \leq 1$. Then,
\begin{equation}\label{eq:lem:compunitary1}
P (\inf_{x \in Comp(\eta,\rho)}  ||H U x||  \leq \gamma\, {C_{\kappa D}}) \leq \epsilon.
\end{equation}
\end{lemma}

{\bf{Proof}:}
We will adopt an argument in the proof of \cite[Lemma 3.3]{RudelsonVershynin_2008}. That is, we will show that the event $E_{c}$ that $|| H U x || \leq \gamma\, C_{\kappa D}$ for some $x \in Comp(\eta,\rho)$,  implies the event $E_{s}$ that $|| H U v || \leq C_{\kappa D} || v||$ for some $v \in Sparse(\eta)$ (for $\rho $ appropriately chosen). 
Note that $P(E_s) \leq \epsilon$ by Lemma~\ref{lem:Bauhutsparseunitary}. If $E_c$ implies $E_s$, then we have $P(E_c) \leq P(E_s) \leq \epsilon$, which is the desired result in  \eqref{eq:lem:compunitary1}.

We first note that every $x \in Comp(\eta,\rho)$ can be written as $x=y+z$, where $v=y/||y||$,   $v \in Sparse(\eta)$ and $||z|| \leq \rho$.  
 Hence  we have the following
\begin{align*}
 || H U  y||  & \leq   ||H U x|| +||H U z|| \\
 		 & \leq   ||H U x|| +||z|| \\
    		& \leq    \gamma C_{\kappa D} +\rho
\end{align*}
where we have used the fact that $||H U z|| \leq ||H U||\, ||z|| \leq ||z||$, and the assumption $||H U x|| \leq \gamma C_{\kappa D}$. Since $||y|| \geq | ||x|| -||z|| | = 1-\rho$,  we can also write the following
  \begin{align}  
 || H U  \frac{y}{||y||} || \leq & \frac{  \gamma \, C_{\kappa D} +\rho}{1-\rho}.
\end{align}
Let us now choose $\rho$ as stated in the condition of the lemma. Then we have  $|| H U  v|| \leq {C_{\kappa D}}$ for some $v \in Sparse(\eta)$, $||v||=1$. Hence we have shown that the event  $E_{c}$ implies the event $E_{s}$. This proves the claim in \eqref{eq:lem:compunitary1}.  
\myqed


We have now established a lower bound for $\inf_{x \in Comp(\eta,\rho)}  ||H U x ||^{2}$ that holds with high probability. 
We now turn our attention to incompressible vectors. For this purpose, we  consider \eqref{eqUnitary:decompincomp}.
We note  that none of the entities in this expression is random. We note the following 
\begin{align}
\nonumber
\inf_{x \in Incomp(\eta,\rho)}  ||\Lambda_x^{-1/2} x ||^{2} 
&= \inf_{x \in Incomp(\eta,\rho)} \sum_{i=1}^N {1 \over \lambda_i} |x_i|^2  \\ 
\label{eqnUnitary:incompsum}
& \geq   \sum_{i \in \psi} {1 \over \lambda_i}   \frac{\rho^2}{2 N},
\end{align}
where the inequality is due to Lemma~\ref{IncompSpread}.  We observe that in order  to have this expression sufficiently bounded away from zero, the distribution of $\frac{1}{\lambda_i}$ should be spread enough.

Let us assume that ${\lambda_i} < C_\lambda^{I} \frac{P}{N-D}$, for $i =D+1, \dots, N$, where $C_\lambda^{I} \in (0,1)$.  Let $ 0.5 \rho^2 \eta N = 0.5 \rho^2 \kappa D> D$. Then we have 
\begin{align}
\nonumber
&\inf_{x \in Incomp(\eta,\rho)}  ||\Lambda_x^{-1/2} x ||^{2} \\
\nonumber
& \geq   \sum_{i \in \psi} {1 \over \lambda_i}   \frac{\rho^2}{2 N} \\
\nonumber
& \geq   (|\psi|-D)    \frac{N-D}{C_\lambda^{I} P}  \frac{0.5 \rho^2}{N} \\
\nonumber
& \geq   (0.5 \rho^2 \kappa D  -D)   \frac{0.5 \rho^2}{C_\lambda^{I} }   \frac{N-D}{N} \frac{1}{P}\\
\label{eqUnitary:infLambdax}
& \geq  {C_I} \frac{D}{P},
\end{align}
where we have used $|\psi| \geq 0.5 \rho^2 \kappa D$, and $C_I$ is defined straightforwardly as in \eqref{eqn:CI}.

We will now complete the argument to arrive at $\prob (\inf_{x\in S^{N-1}}  x^\dagger A x \leq {C}  ) \leq \epsilon $, where $C$ is defined as $\min( \frac{1}{\sigma_n^2} (\gamma \, C_{\kappa D})^2+\frac{1}{\lambda_{max}} ,\frac{D}{P} C_I)$, with $\lambda_{max}$ parametrized as $ \lambda_{max} = C_\lambda^{s} \frac{P}{D}$.  By $\eqref{eqUnitary:decompincomp}$   and $\eqref{eqUnitary:infLambdax}$,  we have $\prob (\inf_{x \in Incomp(\eta,\rho)} x^\dagger A x < C_I \frac{D}{P} )  = 0$. By \eqref{eqUnitary:decompcomp}  and Lemma \ref{lem:compunitary},  we have $\prob ( \inf_{x \in Comp(\eta,\rho)}  x^\dagger A x \leq \frac{1}{\sigma_n^2} (\gamma \, C_{\kappa D})^2+\frac{D}{ C_\lambda^{s}  P} ) \leq \epsilon $. The claim of Lemma~{\ref{lemUnitary:rayleighEigenvalue}}  follows from \eqref{eqUnitary:decomp}.


\end{document}